\documentclass[%
 aip,
 amsmath,amssymb,
 reprint,%
]{revtex4-1}

\usepackage{graphicx}%
\usepackage{dcolumn}%
\usepackage{bm}%

\usepackage[utf8]{inputenc}
\usepackage[T1]{fontenc}
\usepackage{mathptmx}

\usepackage[american]{babel}

\usepackage{url} 			%
\usepackage{color}
\usepackage{hyperref}

\usepackage[outercaption]{sidecap}

\newcommand{\outside}[1][]{_{#1\text{out}}}
\newcommand{\inside}[1][]{_{#1\text{in}}}

\newcommand{\figdir}{figs_150dpi}

\begin{document}
\preprint{arXiv: 1912.06275 [q-bio.BM]}

\title{General Principles of Secondary Active Transporter Function}
\author{Oliver Beckstein}
\email{oliver.beckstein@asu.edu}
\affiliation{Department of Physics, Arizona State University, Tempe AZ 85287, USA}
\author{Fiona Naughton}
\affiliation{Department of Physics, Arizona State University, Tempe AZ 85287, USA}
\affiliation{Present address: Cardiovascular Research Institute, University of California San Franciso, San Francisco, CA 94158, USA}
\date{\today}

\begin{abstract}
  Transport of ions and small molecules across the cell membrane against electrochemical gradients is catalyzed by integral membrane proteins that use a source of free energy to drive the energetically uphill flux of the transported substrate.
  Secondary active transporters couple the spontaneous influx of a ``driving'' ion such as Na\textsuperscript{+} or H\textsuperscript{+} to the flux of the substrate.
  The thermodynamics of such cyclical non-equilibrium systems are well understood and recent work has focused on the molecular mechanism of secondary active transport.
  The fact that these transporters change their conformation between an inward-facing and outward-facing conformation in a cyclical fashion, called the alternating access model, is broadly recognized as the molecular framework in which to describe transporter function.
  However, only with the advent of high resolution crystal structures and detailed computer simulations has it become possible to recognize common molecular-level principles between disparate transporter families.
  Inverted repeat symmetry in secondary active transporters has shed light on how protein structures can encode a bi-stable two-state system.
  Based on structural data, three broad classes of alternating access transitions have been described as rocker-switch, rocking-bundle, and elevator mechanisms.
  More detailed analysis indicates that transporters can be understood as gated pores with at least two coupled gates.
  These gates are not just a convenient cartoon element to illustrate a putative mechanism but map to distinct parts of the transporter protein.
  Enumerating all distinct gate states naturally includes occluded states in the alternating access picture and also suggests what kind of protein conformations might be observable.
  By connecting the possible conformational states and ion/substrate bound states in a kinetic model, a unified picture emerges in which symporter, antiporter, and uniporter function are extremes in a continuum of functionality.
  As usual with biological systems, few principles and rules are absolute and exceptions are discussed as well as how biological complexity may be integrated in quantitative kinetic models that may provide a bridge from structure to function.
\end{abstract}

\maketitle

\vspace{2in}


\tableofcontents

\section{Introduction}
\label{sec:introduction}

Active transporters are integral membrane proteins that move substrate through the membrane against an electrochemical gradient by using a source of free energy.
They are broadly classified as primary and secondary active transporters, depending on the free energy source \citep{Mitchell:1967dx}.
\emph{Primary active transporters} harness chemical reactions (e.g., phosphorylation by ATP) or light.
Some examples are the sodium-potassium pump (Na/K ATPase) \citep{Morth:2007kx}, the rotary F\textsubscript{0}F\textsubscript{1}-ATPase and the light-driven proton pump bacteriorhodopsin \citep{Buch-Pedersen:2009cr}, complex I in the respiratory chain \citep{Sazanov:2015kc}, or ATP-driven ABC transporters such as p-glycoprotein \citep{Lespine:2009lh}.

\begin{figure*}
  \centering
  \includegraphics[width=\textwidth]{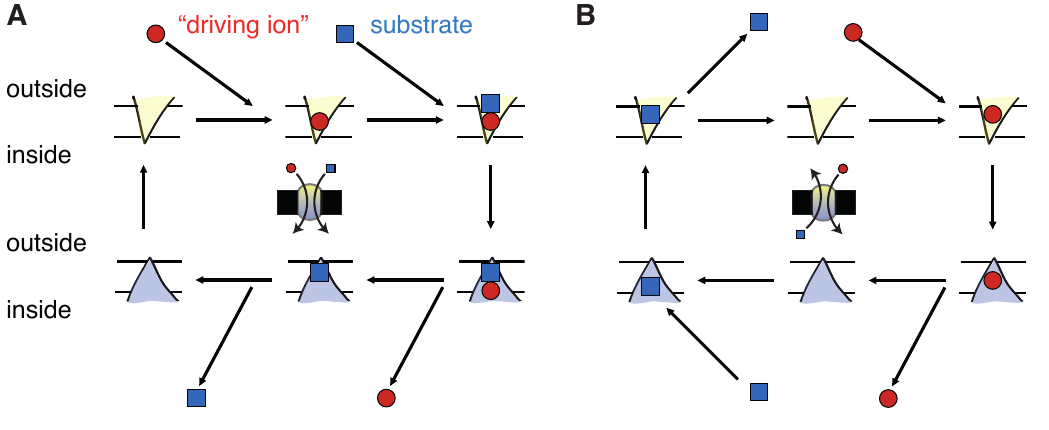}
  \caption[Transport cycle]{Schematic transport cycle of \textbf{A} a symporter (transport of substrate and driving ion in the same direction) and \textbf{B} an antiporter (transport in opposite directions). The central cartoon summarizes the physiological function. The  $\vee$-shaped triangle symbolizes a membrane-embedded transporter protein in the outward facing open conformation in which its binding sites are accessible from the outside. The $\wedge$-shaped triangle indicates the inward facing open conformation. The driving ion is drawn as a red filled circle while the transported substrate is shown as a blue square. The predominant direction of reactions is shown by arrows, with horizontal arrows indicating binding/unbinding and vertical arrows conformational transitions. The order of binding and unbinding events and the stoichiometry of substrate to driving ions may differ from this cartoon.}
  \label{fig:transportcycle}
\end{figure*}

\emph{Secondary transport} is driven by an electrochemical gradient in a \emph{driving ion}, namely sodium or protons.
A few well-studied examples are neurotransmitter transporters (e.g., the serotonin transporter SERT and the dopamine transporter DAT)\citep{Navratna:2019aa}, sodium-proton exchangers (NHE) \citep{Fuster:2014tg}, the calcium exchanger \citep{Ottolia:2007dq}, AE1, the anion exchanger in red blood cells also known as Band 3 \citep{Vastermark:2014kl}, and the divalent anion sodium symporters (DASS) proteins \citep{Sauer:2021ab}.
Secondary active transporters can be broadly divided into two classes based on their physiological behavior \citep{Mitchell:1967dx}.
\emph{Symporters} move their substrate in parallel with the driving ion (Figure~\ref{fig:transportcycle}A).
Both driving ion and substrate are bound at the same time and move in the same direction, typically from the outside to the inside (although ultimately the directionality is determined by the direction and strengths of the gradients), during one half-cycle.
The other half of the transport cycle consists of the movement of the substrate- and ion-free (apo) transporter. 
In the \emph{antiporter} transport cycle (Figure~\ref{fig:transportcycle}B), the driving ion is bound during one half-cycle while in the other half-cycle the substrate is bound and transported in the opposite direction.

Variations of the above scheme are common, though.
For instance, many symporters transport another ion back instead of the apo transition of the transport cycle; for instance, SERT symports serotonin with Na$^{+}$ (and Cl$^{-}$) and counter transports K$^{+}$ (or H$^{+}$) \citep{Keyes:1982aa}.
Sometimes, the driving ion is effectively part  of the substrate as in the AdiC transporter \citep{Fang:2009ye}, which exchanges L-arginine with its decarboxylated product agmatine to effectively export protons.

A third class of related transporters consists of non-coupled transporters. These \emph{uniporters} facilitate diffusion through the membrane. Although we specifically focus on active transporters, the discussion on transport cycles (Section~\ref{sec:unified}) will make clear that the uniporters are closely related to active transporters and it is plausible that small changes in the protein may convert between the two.

\begin{figure*}
  \centering
  \includegraphics[width=\textwidth]{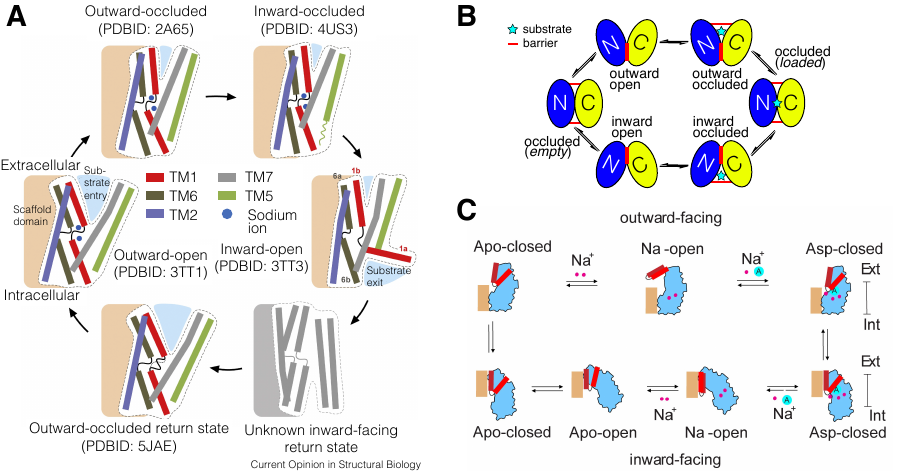}
  \caption{Recent models of symporter transport cycles, based on structural data. \textbf{A} In the bacterial homolog of neuronal sodium symporters, LeuT, a \emph{rocking bundle} movement of the bundle domain (helices TM1, TM2, TM5, TM6, and TM7) relative to the scaffold domain (light brown) enables sodium ions and substrate (not shown) to access the binding sites through water-accessible openings (shaded light blue areas).  (Reproduced with permission from Curr.\ Opin.\ Struct.\ Biol., 54, V.\ Navratna and E.\ Gouaux\protect\citep{Navratna:2019aa}, 161--170 (2019). Copyright 2019 Elsevier.) \textbf{B} Generic cycle of MFS (Major Facilitator Superfamily) transporters (after \protect\citet{Drew:2021aa}). \textit{N} and \textit{C} indicate the N-terminal and C-terminal domain that undergo a \emph{rocker switch} motion and thereby allow the transported substrate (cyan star) to enter and exit the binding site.  \textbf{C} Bacterial homolog of neuronal glutamate transporter, Glt\textsubscript{Ph}, with an \emph{elevator mechanism} whereby the blue transport domain moves between an up and a down position in the membrane, relative to the trimerization domain (wheat) while the reentrant hairpin HP2 (red) governs access of substrate (aspartate, A, cyan circle) and ions (red circles) to the binding site located primarily in the transport domain. (Reproduced with permission from X.\ Wang and O.\ Boudker\protect\citep{Wang:2020aa}, eLife, 9, e58417 (2020). Copyright 2020 the Author(s), licensed under the Creative Commons Attribution 4.0 International (CC BY 4.0) license.)}
  \label{fig:realcyles}
\end{figure*}

The schematic in Figure~\ref{fig:transportcycle} represents a radically simplified view of a transport cycle. In order to provide a sense of the level of simplification, we may compare the symporter cycle (Figure~\ref{fig:transportcycle}A) to recent models for three different classes of symporters shown in Figure~\ref{fig:realcyles}. All three models are based on atomic resolution structures in all major conformations of the cycle in conjunction with experimental functional measurements and often computer simulations. They concisely summarize at a high level current best understanding for three different transport mechanisms \citep{Drew:2016aa}.  For example, LeuT, a sodium-driven symporter for hydrophobic amino acids, progresses through multiple states by a so-called ``\emph{rocking bundle}'' motion whereby one mobile domain moves relative to another static domain \citep{Navratna:2019aa} (Figure~\ref{fig:realcyles}A). In contrast, in Figure~\ref{fig:transportcycle}, the transporter is reduced to a $\vee$ (``vee'') or $\wedge$ (``wedge'') shape, without any regard to the details of the actual molecular organization of the protein.  The transport cycle of Major Facilitator Superfamily (MFS) transporters\citep{Drew:2021aa} (Figure~\ref{fig:realcyles}B) with a symmetrical ``\emph{rocker switch}'' motion between the two structurally similar N and C-terminal domains is broadly reflected in the simple schematic, mainly because the symmetry of the $\vee$/$\wedge$ shapes matches the approximate N/C symmetry. But important details such as the possibility that the binding sites may shift during the cycle and the existence of occluded states, in which the binding sites are not accessible from either compartment, are missing from Figure~\ref{fig:transportcycle}. The so-called ``\emph{elevator}'' transporters (as an example, Figure~\ref{fig:realcyles}C shows the aspartate/sodium symporter Glt\textsubscript{Ph} \citep{Wang:2020aa}) consist of a transport domain, which contains the binding site, and an oligomerization domain, which anchors the protein in the membrane. The transport domain moves up and down through the membrane and thereby switches access to the binding site. The schematic in Figure~\ref{fig:transportcycle} omits the movement of parts of the protein relative to the membrane. It does, however, retain common elements that appear in all three more detailed cycles: Substrate and driving ion bind at the inside of the transporter and a conformational change between \emph{outward facing} ($\vee$) and \emph{inward facing} ($\wedge$) conformations takes place, which ensures that the binding site can only be accessed from either the outside or the inside, known as the \emph{alternating access model}.
It should be kept in mind that despite the additional level of detail in the cartoons in Figure~\ref{fig:realcyles}, important aspects of macromolecular transitions remain oversimplified.
For example, proteins are more flexible and dynamic than suggested by depictions of individual structures with rigid domains and may undergo more gradual conformational changes than the rigid body movements often implied by such cartoons. Furthermore, multiple conformational pathways may have similar probability or pathways may depend on external parameters and thus a simple cycle may be leaving out important information about the actual molecular process.

In this review we focus on overarching principles that are common across many secondary active transporters. The alternating access model provides the ``standard model'' for explaining transporter function in a structural context (Section~\ref{sec:alternatingaccess}). Although evolution always finds ways to add a few exceptions to common rules (for instance, there are a few transporters that do not appear to follow the classical alternating access model), the physical principles under which transporters operate are not negotiable. Transporters function out of equilibrium as ``physical enzymes'' that catalyze transport by free energy transduction through cyclic processes (Section~\ref{sec:thermodynamics}). Ten years ago, a remarkable insight was found into the evolutionary mechanism that can generate protein structures that can switch between the two states of the alternating access model: transporters contain so-called inverted repeat sequences that fold into structures with an internal pseudo two-fold symmetry. This symmetry is broken to generate two different conformations, as discussed in Section~\ref{sec:symmetry}. A complementary view of transporters is that of a pore with multiple coupled gates; originally motivated by the description of ion channels as pores with a single gate, this cartoon model has proven valuable because transporters actually contain physical components that perform the functions of gates, as will be shown with selected examples in Section~\ref{sec:gates}. When the alternating access/gated pore model is considered together with the cycle view of transport, a simple unified picture emerges that describes symporters, antiporters, and uniporters as ideals in a spectrum of functionality (Section~\ref{sec:unified}). We close with a brief perspective in Section~\ref{sec:limitations} on the complexity of observed transporter function that needs to be taken into account beyond the broad principles reviewed here.

\section{The alternating access model}
\label{sec:alternatingaccess}

The \emph{alternating access model} in its basic form was described by \citet{Jardetzky:1966yt} as a polymer molecule that contains binding sites for substrate and is able to assume two different conformations that alternatingly expose the binding sites to the interior and the exterior,  $\vee \rightleftharpoons \wedge$.
The idea of a cyclical process facilitated by a molecule that changes accessibility was expressed by Mitchell in his ``circulating carrier'' model \citep{Mitchell:1957qf, Mitchell:1967dx} and by \citet{Patlak:1957aa} in his gate-type non-carrier mechanism.
Together these models describe in abstract terms a basic framework or model to understand driven transport across the cell membrane.
The key insight was that coupling of two fluxes (substrate and driving ion) could be accomplished by binding to different conformations of the same molecule as discussed in more detail in the next Section~\ref{sec:thermodynamics}.
In particular, it is physically not possible to move substrate against a gradient through a continuous pore, i.e., one that is simultaneously accessible from both sides, regardless of any energy consuming mechanism to open or close the pore \citep{Tanford:1983ek}. The consequence of this insight is that transporters cannot function if continuous pores are formed through the membrane. The alternating access model with its two distinct states provides a conceptual framework that avoids pore formation. However, it requires that a membrane protein is able to change between different conformations on the sub-millisecond timescale\footnote{Turnover numbers of transporters range from one transport event per millisecond (for sodium/proton exchangers) to one per second (some amino acid/cation transporters from thermophiles operating at room temperature) with a typical number on the order of one hundred events per second (e.g., lactose permease); see ``transporter turnover rate'' in the BioNumbers database \url{https://bionumbers.hms.harvard.edu/} \citep{Milo:2010aa}. This means that any step in the transport cycle, including the conformational transition, must be faster than 1 millisecond for the fastest transporters, and 10 milliseconds or 1 second for the slower ones.}, a speed that is easily achievable for macromolecular conformational changes \citep{Henzler-Wildman:2007ri, Schwartz:2009zr, Milo:2010aa}.
The alternating access model also does not give any insights into the actual molecular structure of a transporter except the general requirement that substrate and ion binding sites must switch accessibility in different conformations. In order to obtain deeper mechanistic insights actual atomic-scale structures of transporters in multiple conformations are needed.

\begin{figure}
  \centering
  \includegraphics[width=\columnwidth]{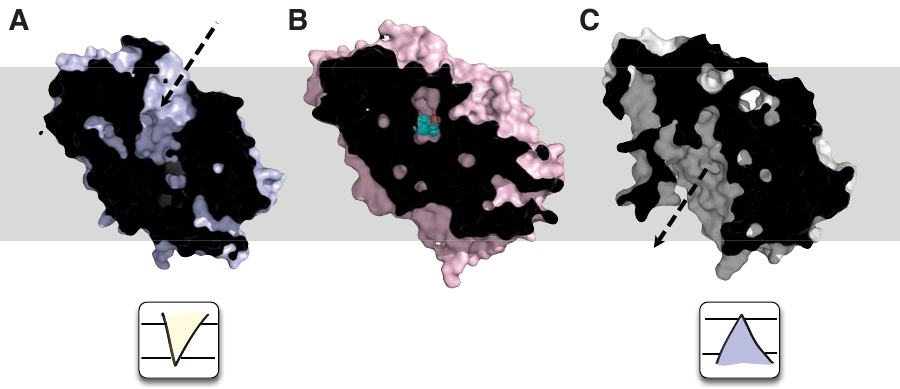}
  \caption{Conformations of the nucleobase/sodium-coupled symporter Mhp1 from X-ray crystallography. \textbf{A} Outward-facing open conformation (PDB ID 2JLN) \citep{Weyand:2008gd}. \textbf{B} Outward-facing occluded conformation with bound substrate benzylhydantoin (PDB ID 4D1B \citep{Simmons:2014pt}; this structure superseded the original 2JLO structure \citep{Weyand:2008gd} but the structural differences are small). \textbf{C} Inward-facing open conformation (PDB ID 2X79; \citep{Shimamura:2010uq}). The approximate position in the membrane is indicated by the gray rectangle in the background. The two cartoons under \textbf{A} and \textbf{C} indicate the two states of the classical alternating access model as used in Figure~\ref{fig:transportcycle}.
    Molecular images were produced with PyMOL \citep{PyMOL}.
  }
  \label{fig:mhp1states}
\end{figure}

The first secondary active transporter for which the major states in the transport cycle were resolved at atomic resolution was the sodium-coupled symporter Mhp1, a member of the nucleobase-cation-symporter 1 (NCS1) family \citep{Weyand:2011bs, Jackson:2013uq}. The structures of wild-type Mhp1 revealed a sodium binding and a substrate binding site deep at the center of the transporter, roughly at the membrane midplane \citep{Weyand:2008gd}. In one structure, these binding sites were accessible from the extracellular side, making this the outward facing (OF, $\vee$) conformation as shown in Figure~\ref{fig:mhp1states}. \citet{Shimamura:2010uq} managed to crystallize wild-type Mhp1 in an inward facing (IF, $\wedge$) conformation in which the binding sites were exposed to the intracellular side. Together they represent the two key conformations required by the alternating access model. A third conformation was also found: in this occluded conformation the binding sites were not accessible from any compartment \citep{Weyand:2008gd, Simmons:2014pt}. The alternating access model does not require such a conformation. As will be argued in Section~\ref{sec:gates}, such occluded states are a necessary consequence of a molecular architecture in which the alternating access conformations are formed by gate domains.

The hallmark of the alternating access mechanism are relatively large conformational changes in the protein conformation and these appear to exist in many secondary transporters for which the alternating access mechanism remains the standard structural framework in which to understand transporter function \citep{Boudker:2010ys, Law:2008ny, Forrest:2009hw, Gouaux:2009zr, Krishnamurthy:2009ij, Abramson:2009ci, Boudker:2010ys, Kaback:2011kx, Forrest:2011ys, Schweikhard:2012uq, Henzler-Wildman:2012ys, Yan:2015aa, Shi:2013qf, Slotboom:2014cr, Diallinas:2014bs, Li:2015fj, Drew:2016aa, Bai:2017aa, Kazmier:2017aa, Henderson:2019aa, Drew:2021aa}.

Although the alternating access model is the canonical model for active transporters, some transporters appear not to be described well within this framework. For example, chloride/proton antiporters are currently understood to function by small changes in a central glutamate residue that alternatingly binds chloride and protons \citep{Miller:2006az, Accardi:2015aa} and do not require the large conformational change---the ``alternating access transition''---that is generally taken to be a key step in the classical alternating access model. However, it could be argued that the fundamental principle of alternating access and the need to maintain a pathway that can not directly conduct ions and substrate always has to be maintained in order to support the cyclical reactions that are required for energy transduction (see the next Section \ref{sec:thermodynamics}) even though different models are also sometimes discussed \citep{Klingenberg:2007uq, Naftalin:2010uq}.

\section{Thermodynamics and cycles}
\label{sec:thermodynamics}

Transport is driven by spontaneous influx of a driving ion. The free energy dissipation from flowing down the driving ion's electrochemical gradient is coupled to the vectorial transport of a substrate molecule or ion.
\citet{Hill:1989ve} clearly explained the principle of free energy transduction in transporters (and enzymes) through 
a cyclic process that tightly couples driving ion flux and substrate flux.
Following his treatment, we will first qualitatively explain how a cyclical process that operates out of equilibrium transduces energy. We will then briefly revisit the thermodynamic driving forces of the process in order to motivate the idea that transporters are enzymes that catalyze transport.

\subsection{Transport is a non-equilibrium process}
\label{sec:noneq}

\begin{figure*}
  \centering
  \includegraphics[width=\textwidth]{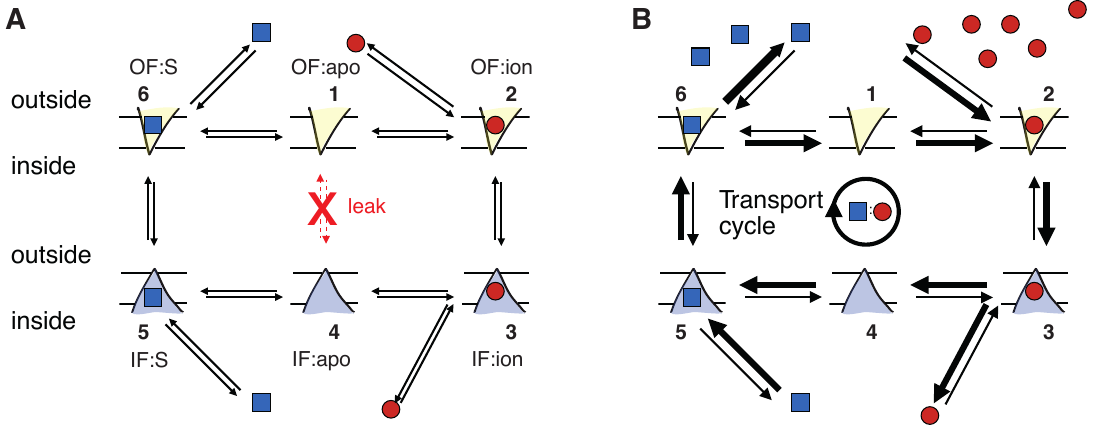}
  \caption{Transport by an antiporter is a cyclical out-of-equilibrium process. The simplified cycle only includes outward facing open (OF) and inward facing open (IF) states with either substrate (S, blue square) or driving ion (ion, red filled circle) bound or nothing bound (apo). \textbf{A} Equilibrium---all concentrations are at equilibrium values and all reactions obey detailed balance. A potential leak pathway $1 \rightleftharpoons 4$ (red dashed arrows) is forbidden in the ideal cycle.  \textbf{B} Out-of-equilibrium---the outside ion concentration is raised over its equilibrium value, which leads to moving all states out of equilibrium. The cycle is driven in the clockwise direction, resulting in stoichiometric 1:1 transport. The states are numbered so that one can refer to, say, the outward facing apo state (neither ion nor substrate bound) of the transporter as $T_{1}$ or the inward facing, substrate-bound state as $T_{5}:S$ where the presence of the substrate is included for clarity even though the label ``5'' includes the presence of the substrate (as opposed to state 4, which does not include it).}
  \label{fig:antiportercycle}
\end{figure*}

Consider, for instance, a hypothetical antiporter that uses one driving ion (red circle in Figure~\ref{fig:antiportercycle}A) to move one substrate molecule (blue square).
We initially imagine the system to exist in equilibrium, i.e., the net fluxes between all states are zero, also known as the detailed balance condition. The inside and outside populations of particles $i$ are in Nernst equilibrium, i.e., when considering the concentrations on either side of the membrane and the membrane potential, no net flux of particles would occur if a pore selective for species $i$ were opened in the membrane.\footnote{It is also necessary that the transporter state populations are in equilibrium with each other and the ion and substrate concentrations. However, any imbalance in the transporter populations would soon move towards the equilibrium values, provided that the ion and substrate concentrations are at equilibrium. A transport cycle cannot be driven by the transporter; only the binding/dissociation of external ions and substrates can continuously draw on a source of free energy.}
For example, the binding of a driving ion $I$ to the empty transporter in the outward facing conformation $T_{1}$ is the equilibrium reaction
\begin{gather}
  T_{1} + I \rightleftharpoons T_{2}:I.
  \label{eq:eqT1T2}  
\end{gather}
where $T_{2}$ represents the outward-facing, ion-bound state.
The isomerization between outward facing and inward facing conformation (the alternating access transition) is
\begin{gather}
  T_{2}:I \rightleftharpoons T_{3}:I.
  \label{eq:eqT2T3}
\end{gather}
where $T_{3}$ represents the inward-facing, ion-bound state.
(Similarly, state $T_{4}$ stands for the inward-facing apo transporter, $T_{5}$ is the inward-facing transporter with substrate bound, and $T_{6}$ is the outward-facing, substrate-bound state.)
Because all individual fluxes are zero, no net transport takes place. On average, for every substrate molecule that is moved from inside to outside in a given unit of time, the same number of molecules are moved from the outside to the inside.
We now perturb the system away from equilibrium by increasing the outside concentration of the driving ion, as indicated by the larger number of driving ions in Figure~\ref{fig:antiportercycle}B. Following Le Chatelier's principle, the equilibrium of the binding reaction Eq.~\ref{eq:eqT1T2} is moved as to increase the concentration of products \citep{MolecDrivingForces03}, i.e., the number of ion-loaded transporters $T_{2}:I$ increases above its equilibrium value. Because the reactants (inputs) of the isomerization reaction Eq.~\ref{eq:eqT2T3} are provided by the products (output) of the binding reaction Eq.~\ref{eq:eqT1T2}, which have increased, Le Chatelier's principle equally applies to the isomerization and pushes this equilibrium towards the ion-loaded inward facing conformation, $T_{3}:I$. The same reasoning is applied to each subsequent reaction and in this way, net flux of substrate from the inside to the outside is induced in steps 5$\rightarrow$6. Crucially, the reactions form a \emph{cycle} so that after the steps 1$\rightarrow$2$\rightarrow$3$\rightarrow$4$\rightarrow$5$\rightarrow$6$\rightarrow$1 the transporter is in exactly the same state as it was before. However, the environment has changed as one ion was transported from the outside to the inside and one substrate was transported from the inside to the outside with 1:1 stoichiometry, as expressed by the transport reaction of the antiporter
\begin{gather}
  \label{eq:antiporter}
  I\outside + S\inside \rightarrow I\inside + S\outside
\end{gather}
where $I$ stands for the driving ion and $S$ for the substrate and subscripts indicate their location relative to the membrane.
The corresponding reaction of the symporter is
\begin{gather}
  \label{eq:symporter}
  I\outside + S\outside \rightarrow I\inside + S\inside.
\end{gather}
Different stoichiometries require different stoichiometric coefficients. For instance, a 2:1 antiporter would be described with $2I\outside + S\inside \rightarrow 2I\inside + S\outside$.
\citet{Hill:1989ve} makes the above reasoning fully quantitative by considering how the net fluxes between states, which are zero in equilibrium, become biased in one direction when a component is perturbed.
The resulting theory of cycle fluxes can be applied to arbitrarily complex cycles to analytically compute steady state populations and fluxes.

More realistic transporter schemes contain additional transitions such as the one between the two apo states $1 \rightleftharpoons 4$, often referred to as leaks or slippage. Such a transition would allow three cycles to become possible: The transport cycle that was just described and two leak cycles: the ion leak cycle $1 \rightleftharpoons 2 \rightleftharpoons 3 \rightleftharpoons 4 \rightleftharpoons 1$ and the substrate leak cycle $1 \rightleftharpoons 6 \rightleftharpoons 5 \rightleftharpoons 4 \rightleftharpoons 1$. Under physiological conditions, the ion leak cycle would dissipate the ionic gradient. Cells spend a substantial amount of their chemical energy to establish the driving ion gradient. In mammals an estimated 19\%--28\% of ATP are used to power the Na\textsuperscript{+}-K\textsuperscript{+}-ATPase that establishes the transmembrane sodium gradient \citep{Rolfe:1997pe}. Therefore, dissipation of the sodium gradient is costly and reduces the organism's fitness. The substrate leak cycle would run in the opposite direction and let substrate molecules enter the cell, counter to the physiological necessity of the transporter to remove them from the cell. Under physiological conditions, leak cycles must be suppressed by decreasing the rate for slippage transitions such as $1 \rightleftharpoons 4$.

The qualitative discussion makes clear that energy transduction, i.e., the use of the free energy stored in the driving ion gradient, requires a \emph{complete cycle} that contains both ion and substrate translocation steps. If any part of the cycle is broken, no energy transduction can take place. Thus, energy transduction is a property of complete cycles and not of individual states \citep{Hill:1989ve}. Therefore, there is no specific step in the cycle that could be described as an ``energized'' state or a state where energy is ``gained by a binding reaction'' \citep{Hill:1981aa}.

In general, a protein that functions according to the alternating access mechanism cannot function if it presents a continuous, leaky pathway \citep{Tanford:1983ek} as this prevents energy coupling. Similarly, nonproductive leak cycles also reduce the efficiency of a transporter. Although here we generally discuss ideal, fully efficient cycles to elucidate the basic principles, real transporters leak and therefore their transport stoichiometry is generally not the ideal one \citep{Hill:1989ve, Henderson:2019aa}. For example, instead of an ideal 1:1 stoichiometry one might measure only 1:0.75, i.e., on average 1.33 driving ions are needed to move one substrate because only 75\% of the total flux comes from productive cycles (1:1 stoichiometry) and 25\% comes from leak cycles (1:0).
However, not all leak cycles need to be deleterious for an organism.  Recent work by Zuckerman and collaborators (further discussed in Section \ref{sec:unified}) indicates that leak cycles enable kinetic control over selectivity \citep{Bisignano:2020aa}. The rates for these slippage transitions could serve as an additional set of molecular parameters that may be varied by evolution to increase organismal fitness.

The ion and substrate binding or dissociation steps are necessary components of the cycle because without them the cycle cannot be driven in a specific direction: these steps provide the only external ``handle'' to control the process \citep{PLoC:2019ab}. Therefore, no cyclical process with a net flux in one direction exists in which only a protein changes through a repeated sequence of conformational states; coupling to an external source of free energy is always necessary.

Finally, it is worth emphasizing that because the transporter protein moves cyclically through different conformations, it is not altered in any permanent way. In the energetic description of the process (see Section~\ref{sec:drivingforces} below), the transporter does not appear. Thus, transporters act as enzymes for moving substrate, similar to how biochemical enzymes catalyze the formation and breaking of chemical bonds. In this sense, transporters are ``physical enzymes'' or ``molecular machines'' in that they catalyze a physical process instead of a chemical one. Other proteins of this kind are molecular motors, which turn chemical energy into movement of the protein itself, or rotary pumps such as the V-type and P-type ATPases, which turn chemical energy into rotary motion and movement of protons or ions across the cell membrane; the latter can run in reverse to turn rotary motion by ion flow into chemical bonds. Similarly, transporters run backwards if the electrochemical gradients are changed appropriately, as experimentally demonstrated in a wide range of different transporters (chloride-proton transporter ClC-7 \citep{Graves:2008aa}, proton-coupled peptide transporter PepT\textsubscript{St} \citep{Parker:2014qf}, bacterial Na$^{+}$-coupled succinate transporter VcINDY \citep{Mulligan:2014aa, Fitzgerald:2017aa}, bacterial sodium/sugar symporter vSGLT \citep{Fitzgerald:2017aa}). An analysis of the thermodynamic driving forces for the cycle in the following section will show that the external electrochemical gradients (concentrations and membrane potential) are the \emph{only} factors determining the direction of transport. However, thermodynamics does not determine kinetics and hence it is possible that the speed of the cycle differs in forward and reverse direction. In particular for any realistic transport cycle, switching the driving forces will drive the cycle in the opposite direction but the turnover number in the reverse direction can be very different from that in the forward direction, as, for example, observed for the glutamate transporter EAAC1, where reverse glutamate transport was faster than forward transport \citep{Zhang:2007ab}.

\subsection{Driving forces}
\label{sec:drivingforces}

Quantitatively, the only thermodynamic driving forces $X_{i}$ are the ones originating in electrochemical potential ($\mu' = \mu_{0} + kT\ln c/c_{0} + q\Psi$) differences of ions and substrates across the cell membrane \citep{MolecDrivingForces03}; free energy differences due to the different states of the protein cancel in the whole cycle and play no role \citep{Hill:1989ve}. The driving force for species $i \in \{I, S\}$ is
\begin{align}
  \label{eq:drivingforces}
  X_{i} = \mu'\inside[i,] - \mu'\outside[i,] = kT\ln\frac{c\inside[i,]}{c\outside[i,]} + q_{i}\Delta\Psi
\end{align}
where  $c_{i}$ is the concentration (or activity) on the indicated side of the membrane, $q_{i}$ the charge, $\Delta\Psi = \Psi\inside - \Psi\outside$ is the transmembrane potential, $T$ is the temperature and $k$ is Boltzmann's constant.
The membrane potential is typically negative, $\Delta\Psi < 0$. Thus, for typical driving cations (Na$^{+}$, H$^{+}$ with $q=+1e$) and $\Delta\Psi \approx -100\,\text{mV}$ the membrane potential contributes at $T=298$~K about $q_{I}\Delta\Psi \approx -3.9\,kT$. 
Typical sodium concentrations are on the order of $100$~mM on the outside and $10$~mM inside a cell and hence $kT\ln\frac{c\inside[I,]}{c\outside[I,]} = -2.3\,kT$.  If the substrate is neutral (the electrostatic component is zero for $q_{S}=0$) then a positive net charge is moved into the cell down an electrostatic potential and a sizable fraction of the available free energy will be provided by the membrane potential component. In general, any electrogenic transport (movement of a net charge) is affected by the membrane potential.

Denote by $J_{i}$ the flux at which particle $i$ is transported across the membrane (in particles per unit time), with the direction out$\rightarrow$in counting as $J_{i} < 0$ and the reverse as $J_{i} > 0$. Note that in a simple cycle such as the one in Figure~\ref{fig:transportcycle}, exactly one ion is moved for each substrate molecule and hence the absolute values of these fluxes must be the same, $|J_{I}| = |J_{S}|$ but the signs will differ, depending on symport or antiport processes.

When the driving force is negative, e.g., $X_{I} < 0$, then spontaneous movement occurs, such as influx of the driving ion and hence $J_{I} < 0$. The antiporter is supposed to move substrate against a gradient from the inside to the outside, i.e., against the opposing driving force $X_{S} < 0$ under which $S$ particles would spontaneously move into the cell. The rate of free energy dissipation is
\begin{gather}
  \label{eq:dissipation}
  \Phi = J_{I}X_{I} + J_{S}X_{S} \ge 0.
\end{gather}
$\Phi = 0$ holds in equilibrium but then no transport occurs (see Section~\ref{sec:noneq}). The second law of thermodynamics requires $\Phi > 0$ in non-equilibrium steady state, i.e., when concentrations remain fixed at their non-equilibrium values and do not change \citep{Hill:1989ve}. In steady-state, the transporter moves ions and substrates at a constant flux. Under which conditions will the antiporter move $S$ from inside to outside, i.e., given $J_{S} > 0$ (even though $X_{S} < 0$), what is required of $I$? Rearranging Eq.~\ref{eq:dissipation}
\begin{gather}
  \label{eq:JS}
  J_{I} X_{I} > -J_{S} X_{S}
\end{gather}
and noting that the right-hand side is positive, it follows that $J_{I} X_{I}$ also has to be positive, i.e., the driving ion must flow down its electrochemical gradient from the outside to the inside ($J_{I}<0$, $X_{I}<0$). In other words, spontaneous fluxes always dissipate free energy, which can be coupled to the substrate flux. This free energy dissipation rate must be larger than the rate of free energy required to move $S$ against its driving force. For a simple antiporter cycle without leakage, $J_{S} = -J_{I}$ (for each $I$ transported to the inside, one $S$ is transported to the outside, in the same amount of time) and hence $-J_{S} X_{I} > -J_{S} X_{S}$ and with $J_{S}>0$,
\begin{gather}
  \label{eq:XI}
  X_{I} < X_{S} \quad\text{(simple antiporter)}
\end{gather}
is required for transport, i.e., the electrochemical gradient of the driving ion must be steeper and more negative than that of the substrate. In other words, the amount of available free energy per driving ion translocation event must be larger (more negative) than the substrate gradient against which $S$ is moved because out of equilibrium not all free energy can be transformed into useful work and a fraction always increases the entropy of the universe in the form of heat, as required by the second law. The condition Eq.~\ref{eq:JS} can also be fulfilled with $J_{I}>0$, $X_{I}>0$, i.e., a spontaneous flux of ions from the inside to the outside. In this case the transporter would need to operate as a symporter to move driving ion and substrate \emph{together}---energetic coupling cannot happen in separate cycles \citep{Hill:1989ve}).

For a simple symporter with $X_{I}< 0$, $X_{S} > 0$ and $J_{I} = J_{S} < 0$, the condition equivalent to Eq.~\ref{eq:XI} reads
\begin{gather}
  \label{eq:XIsymp}
  X_{I} < -X_{S} \quad\text{(simple symporter)}.
\end{gather}
We will come back to the question of the relationship between symporters and antiporters in Section~\ref{sec:unified} where we will see that one can write a universal kinetic scheme that encompasses symporters, antiporters, and uniporters.

\section{Inverted repeat symmetry}
\label{sec:symmetry}

The alternating access model together with the thermodynamic cycle analysis explains how transporters function in principle, i.e., they describe the physical constraints under which any transporter protein has to operate.
However, understanding \emph{how} these principles are embodied in an actual biomolecule requires structural atomic-resolution data, primarily provided by X-ray crystallography and electron microscopy.
The most important requirement of the alternating access model is the existence of \emph{two states} that make binding sites accessible to the outside \emph{or} the inside, generally referred to as an \emph{outward facing (OF) conformation} and an \emph{inward facing (IF) conformation}. It has been hypothesized that a viable evolutionary path to create a switchable two-state system in a single protein molecule can be based on an internal twofold structural symmetry, so-called \emph{inverted repeats}. This deep insight into a fundamental principle of transporter function was only recently discovered by \citet{Forrest:2008ax} and since then broadly recognized as nearly universal \citep{Forrest:2013fk, Forrest:2015lr}.

\subsection{Inverted repeat structures}
\label{sec:symmfolds}

Structural symmetry is well represented among membrane proteins. These symmetries can arise due to oligomerization, often seen in cyclically symmetric channels and pores, or due to the presence of internal repeats within the protein sequence \citep{Forrest:2015lr}. Notably, internal repeats occur more frequently in membrane protein superfamilies than overall \citep{Myers-Turnbull:2014aa}; in the case of secondary active transporters, most known structures show an \emph{inverted repeat symmetry} \citep{Bai:2017aa,Shi:2013qf}---that is, internal repeats which adopt similar folds but start on opposite sides of the membrane, giving rise to \(C_2\) pseudosymmetry about an axis parallel to the membrane plane. Further common to most secondary active transporters is the presence of two bundles or domains, with the substrate binding sites located near the interface and often involving discontinuous helices \citep{Shi:2013qf}. The exact number of transmembrane helices and distribution of the inverted repeats over the two domains differs, with several common folds observed (Figure~\ref{fig:invertedrepeats}):

\begin{figure*}
  \centering
  \includegraphics[width=\textwidth]{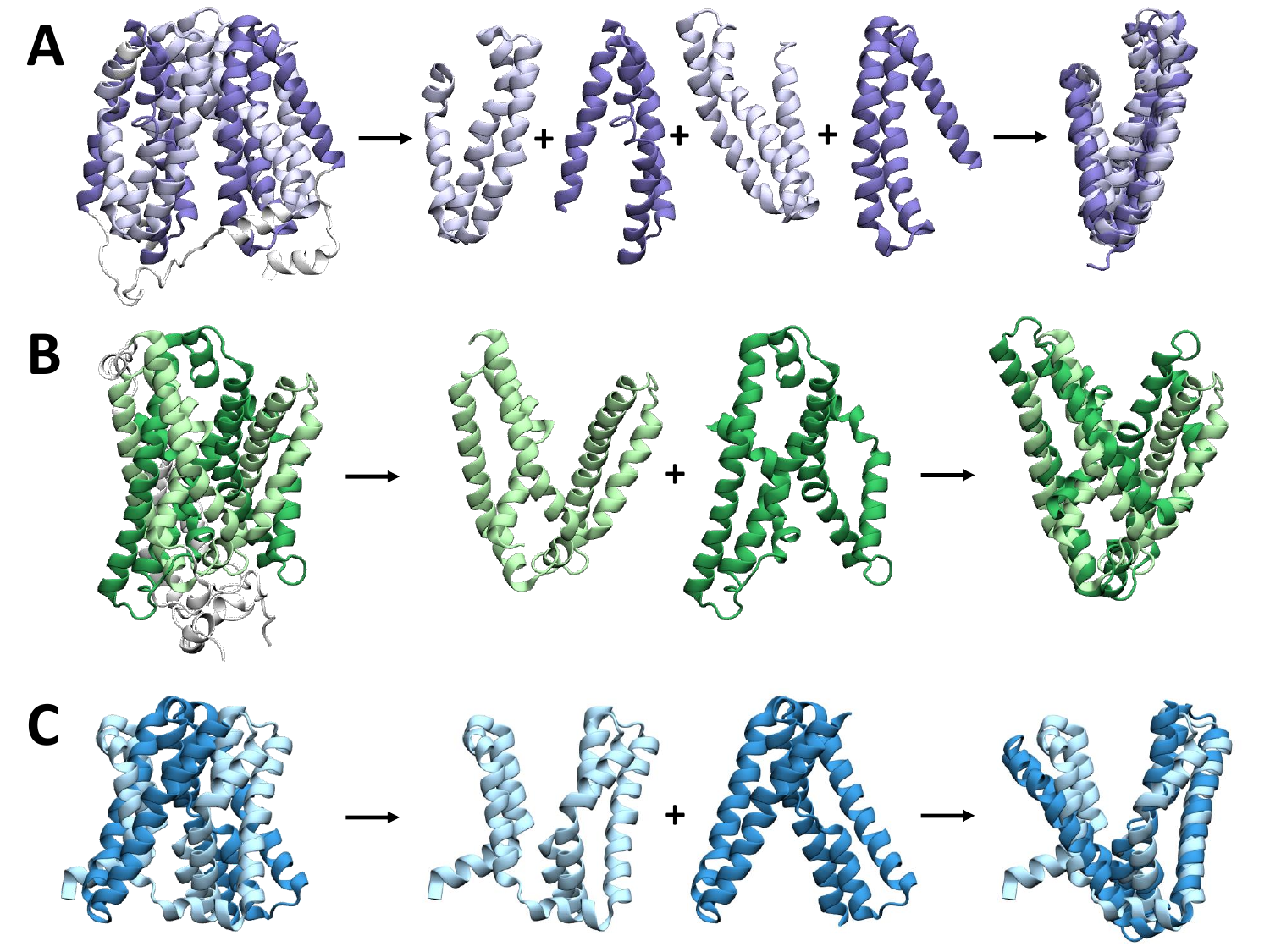}
  \caption{Representative structures of common secondary active transporter folds, highlighting the inverted repeats. In each case, the protein is first shown whole, then with the repeats separated translationally, and finally with the second repeat rotated and overlayed on the first.
    \textbf{A} LacY (inward facing, MFS fold; PDB ID IPV6)
    \textbf{B} Mhp1 (outward facing, LeuT fold; PDB ID 2JLN)
    \textbf{C} ASBT\textsubscript{NM} (inward facing, NhaA fold; PDB ID 3ZUY).
    Molecular images were produced with VMD \citep{Hum96}.
  }
  \label{fig:invertedrepeats}
\end{figure*}

\paragraph{(3 + 3) + (3 + 3): Major Facilitator Superfamily (MFS).}

The core MFS fold contains 12 transmembrane helices (TMs) \citep{Law:2008ny, Yan:2015aa, Drew:2021aa}. An N- and C-domain are each formed from a pair of 3 TM inverted repeats, and are themselves twofold pseudosymmetric. The MFS is one of the largest transporter families found across multiple organisms; the first structures were reported for lactase permease LacY \citep{Abramson:2003fe} and glycerol-3-phosphate transporter GlpT \citep{Huang:2003aa}, with many determined since.

\paragraph{(5 + 5): LeuT fold.}

The LeuT fold consists of 5 TM inverted repeats, with the first two helices from each repeat forming a core bundle  (i.e. TMs 1, 2, 6, 7), while the next two (TMs 3, 4, 8, 9) form a scaffold/hash domain; TMs 5 and 10 may act as gates \citep{Kazmier:2017aa}. First observed in the neurotransmitter/sodium symporter LeuT \citep{Yamashita:2005wv}, several other transporters have been found to adopt this fold, including the sodium/hydantoin transporter Mhp1 \citep{Weyand:2008gd}.

\paragraph{(5 + 5): NhaA fold.}

Also consisting of 5 TM repeats, the NhaA fold is observed in sodium/proton antiporters (e.g. NhaA \citep{Hunte:2005tg}) and the apical sodium dependent bile acid transporter (ASBT) family (e.g. ASBT\textsubscript{NM} \citep{Hu:2011bh}). The first two helices of each repeat (TMs 1, 2, 6, 7) form a panel or dimer domain, with the remaining three (TMs 3-5, 8-10) forming a core domain \citep{Fuster:2014tg, Padan:2014jl}.

\paragraph{(7 + 7): 7-TM inverted repeat (7TMIR) fold.}

Relatively recently identified, transporters with this fold include the proton/uracil symporter UraA and chloride/bicarbonate antiporter AE1; four helices from each repeat (TMs 1-4, 8-11) for a core domain, while the remaining three (TMs 5-7, 12-14) form a gate domain \citep{Chang:2017aa}.

\subsection{Origin and function of inverted repeats}
\label{sec:originfunctioninvertedrepeats}

Internal repeats such as these have been speculated to arise from the duplication of an ancestor gene and subsequent fusion event, in this case following a flip of one duplicate relative to the membrane; possible candidates showing these initial ``half'' folds have been identified in the DedA (for the LeuT fold) and SWEET (for the MFS fold) families \citep{Keller:2014ly}. The EmrE multidrug transporter is proposed to come together as an antiparallel dimer and function through an exchange of asymmetrical structures similar to that described below \citep{Korkhov:2009fk,Morrison:2012fk}, and represents a possible pre-fusion step in the proposed duplication-and-fusion evolutionary process of inverted repeat symmetry.

Distinct inward- and outward-facing conformations arise from asymmetry in the exact conformations of the repeats composing the two domains (discussed in Section \ref{sec:asymmetry}), which changes the relative locations/orientations of these domains. Several mechanisms for this relative motion have been proposed \citep{Drew:2016aa,Forrest:2011ys}: the domains may rotate about the substrate binding site to alternatively expose it to each side of the membrane, as in the \emph{rocker-switch} (where the domains are structurally symmetric, proposed for MFS transporters \citep{Radestock:2011aj}) and \emph{rocker-bundle} (where the domains are distinct, e.g. for LeuT fold transporters \citep{Kazmier:2017aa}) mechanisms; or, as in the \emph{elevator mechanism}, one domain predominantly containing the binding site may move perpendicular to the membrane, moving the binding site against to relatively fixed second domain to expose it to each side of the membrane in turn (proposed for NhaA fold transporters \citep{Padan:2014jl, Drew:2016aa}).

\subsection{Asymmetry and alternating access}
\label{sec:asymmetry}

\begin{figure*}
  \centering
  \includegraphics[width=\textwidth]{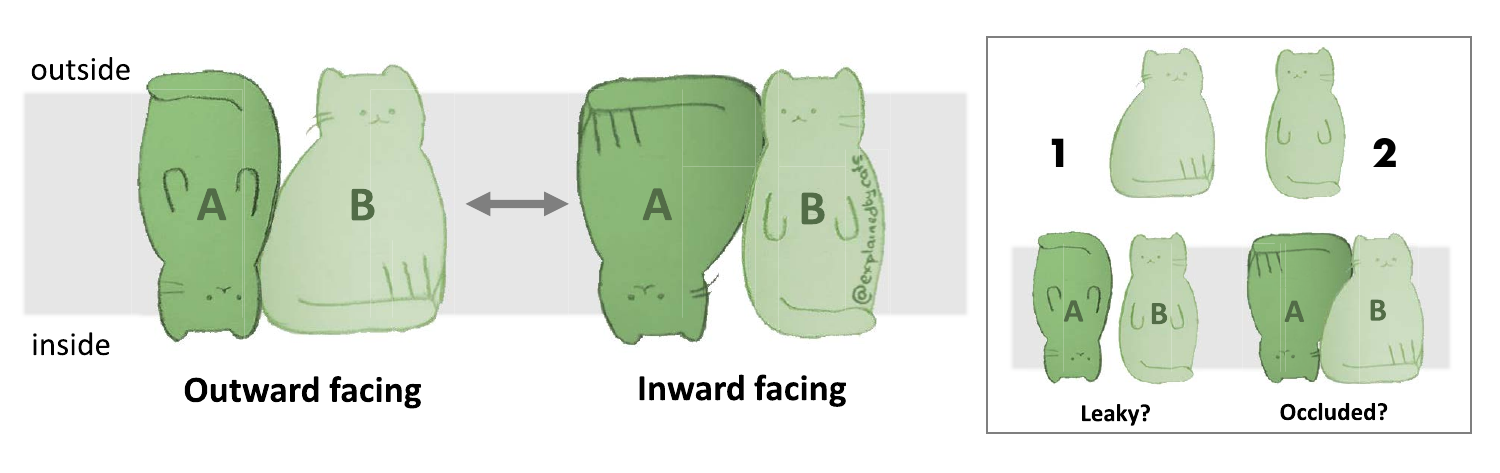}
  \caption{Cartoon showing how symmetry-broken inverted repeats generate the two major conformations in the alternating access mechanism. Each repeat (labeled A and B) may take on one of two conformations (shown as 1 and 2 in the inset), giving rise to 2 x 2 = 4 possible conformations, though two (occluded and ``leak'' states, shown inset) are not part of the basic alternating access mechanism. [Drawings by Fiona Naughton (@explainedbycats).]}
  \label{fig:symmetrybreaking}
\end{figure*}

While inverted repeats share an overall fold, they are found to take on different conformations, giving rise to an asymmetry that allows the substrate binding site to be exposed to one side of the membrane while blocked from the other. By exchanging conformations between the two repeats---with the first repeat adopting the conformation of the second and vice versa---the protein is thus able to switch between an inward facing and an outward facing state \citep{Forrest:2011ys} (Figure~\ref{fig:symmetrybreaking}). By breaking the symmetry between the two inverted repeats in a symmetrical fashion, two conformational states are naturally created.

Exchanging conformations in this way is hypothesized to require no or little net energetic change in the overall protein structure from the inward-to-outward or outward-to-inward transitions \citep{Forrest:2015lr}. The presence of two possible conformations for each of the two repeats also brings up the question of whether the repeats can possess the same conformation at a given time; such overall conformations might form occluded (closed at both side) or leaky (open on both sides) states of the transporter (Figure~\ref{fig:symmetrybreaking}; inset). The presence of occluded and leak states in the transport cycle is discussed further in Section \ref{sec:gates}.

The above described \emph{repeat swapping} has been taken advantage of to generate homology models of transporters in different states, given a structure in only one state: the conformation of each repeat is used as a template for the other, forcing the exchange of conformations \citep{Vergara-Jaque:2015aa}. This method was first applied to LeuT \citep{Forrest:2008ax}, producing a structure that latter proved to be consistent with an experimental structure \citep{Krishnamurthy:2012za}, and has since been used to generate structures for a range of secondary active transporters, with subsequent experimental validation obtained in several cases; including the aspartate transporter Glt\textsubscript{Ph} \citep{Crisman:2009ol, Reyes:2009aa}, LacY \citep{Radestock:2011aj}, CcdA \citep{Zhou:2018aa}, and NhaA \citep{Schushan:2012aa}.

\section{Transporters as gated pores}
\label{sec:gates}

\begin{figure}
  \centering
  \includegraphics[width=\columnwidth]{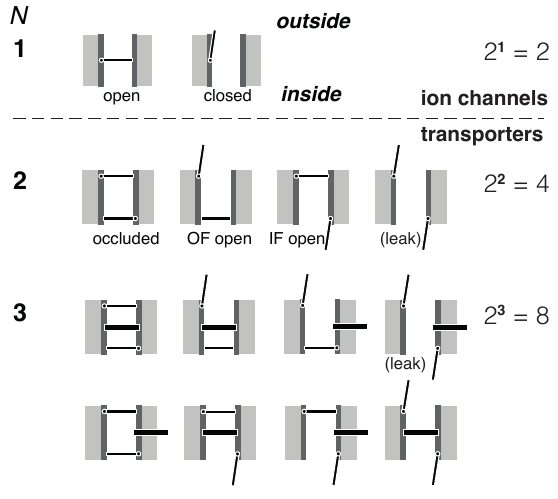}
  \caption{Ion channels and transporters as gated pores. In this simplified picture, ion channels contain a single gate that is controlled by external stimuli. Transporters can implement alternating access by the coordinated movement of two or more gates. The number of gates $N$ determines the total number of distinct states, $2^{N}$. Not all states might be physiologically observed, and some, such as the leak states, will prevent energy transduction.}
  \label{fig:gatestates}
\end{figure}

\citet{Lauger:1980qf} envisaged ion channels as pores with a single free energy barrier, which can be identified with the gate of the channel that controls ion flow in response to external stimuli \citep{Hille01}.
He could model transporters as pores with two coupled barriers \citep{Lauger:1980qf}; motivated by the suggestive
original ``gated pore model'' \citep{Klingenberg:1979aa} (which was more precisely renamed the  ``single binding center gated model'' (SBGP) \citep{Klingenberg:2007uq}) one may also call these two barriers gates (Figure~\ref{fig:gatestates}) \citep{Abramson:2009ci, Krishnamurthy:2009ij, Forrest:2011ys}. Such a gate should be thought of as a switch or bi-stable element that can exist in two states that are generally called ``open'' and ``closed'' although it might also carry the meaning ``outward facing'' or ``inward facing''. The gate picture reduces the nuanced view of free energy barriers with variable barrier height to one in which either a very high barrier exists (``closed'') or the barrier is small compared to thermal fluctuations (``open''). This simplification allows one to broadly enumerate and categorize states and make general (but necessarily approximate) statements for whole classes of proteins. It also allows one to create simple cartoons of transporter states that summarize transporter conformations succinctly.
As will be shown below, the cartoon is a useful abstraction because gates correspond to physical molecular domains in transporter proteins (i.e., they have a molecular identity) and thus states generated from the gate picture directly correspond to observable protein conformations.

In ion channels, a change in membrane potential or binding of a signaling molecule opens the gate and ions spontaneously flow through the open pore down their electrochemical gradient \citep{Hille01}, as shown for a pore with $N=1$ gate in Figure~\ref{fig:gatestates}.\footnote{Describing a channel with a single gate is an oversimplified picture because many ion channels contain an activation gate and a separate inactivation gate, which switches the channel into a relatively long lasting inactive state \citep{Hille01}. These gates are typically coupled to some degree, e.g., the inactivation gate might only close when the activation gate has closed but the coupling is presumably different from the coupling between gates that enables energy transduction in transporters.}
In the ``transporters as gated pores'' picture, the coordinated movement of two (or more) gates creates the conformations of the alternating access model (Section \ref{sec:alternatingaccess}).
The outward facing open state is formed when the outer gate opens while the inner gate remains closed. Conversely, the inward facing state consists of the outer gate closed, while the inner one is open, as shown for a transporter with $N=2$ gates in Figure~\ref{fig:gatestates}.
Simultaneous opening of both gates must be avoided---a ``leak'' state (see Figure~\ref{fig:gatestates})---to prevent leakage of the driving ion and dissipation of the ionic gradient. The coordination of the gates is termed ``coupling''. Furthermore, conformational changes (i.e., changes in the gates) must also be coupled to the binding/dissociation of driving ion(s) and substrate(s), a point that will be revisited below and explicitly included in Section~\ref{sec:unified}.  
For simplicity, we will focus on the different conformational states of the protein while keeping in mind that the presence of ions and/or substrates will likely change the structure to some degree.

It has been observed experimentally that under certain conditions transporters can also function as channels \citep{DeFelice:2007dz}, a view that fits naturally in the picture of a transporter whose gates are not fully coordinated so that leak states may occur (see Figure~\ref{fig:gatestates}). Channel-like behavior is characterized by spontaneous energetically downhill diffusion of ions (or substrates) in a non-stoichiometric and burst-like fashion, which differs from the leak cycles that can occur in the standard cycle due to loose coupling \citep{Henderson:2019aa} in that the latter still only move individual particles. For the aspartate transporter Glt\textsubscript{Ph}, a cryo-electron microscopy structure in combination with MD simulations showed that the long-sought chloride channel conductance pathway \citep{Vandenberg:2008oq} is formed in an intermediate step between the outward facing and the inward facing conformation \citep{Chen:2021ab}.

Nevertheless, the ``transporters as gated pores'' picture is more than just a convenient cartoon model because as we will discuss below, the gates generally represent a molecular reality, i.e., secondary transporters contain distinguishable parts that function as gates. Therefore, conformational states that are predicted from the gated pore model generally correspond to conformations with distinct structural arrangements of the corresponding gates.

\subsection{Gates as molecular building blocks}
\label{sec:entities}

Since X-ray crystallography and cryo-electron microscopy have revealed the molecular structures of a range of transporters, various authors have identified domains of these proteins that regulate access to binding sites with gates, as summarized by \citet{Forrest:2011ys}. A particular terminology of \emph{thin} and \emph{thick} gates originated in the structural analysis of LeuT-like transporters \citep{Krishnamurthy:2009ij, Abramson:2009ci}. Thin gates are generally considered to be parts of the protein whose movement can prevent the exchange of ions or substrates with the intra- or extracellular solution. Perhaps somewhat confusingly, the conformational transition that is responsible for alternating access, or rather the sum of moving structural elements, is sometimes considered the thick gate. In other transporter families, such as the MFS transporters, no special distinction between thin and thick gates is commonly made. 

Although there is some ambiguity in how to define gates, they are nevertheless recognizable molecular entities. \citet{Diallinas:2014bs} concludes, based on work in the purine transporter UapA, that physiological transport properties are determined by intramolecular interactions between binding sites and gating elements, similar to ones present in channels.
\citet{LeVine:2016aa} quantitatively analyzed the mechanism of the LeuT transporter with a particular emphasis on the allosteric coupling between ions, substrate, and the protein. Based on experimental and simulation data, they concluded that LeuT is best described with an \emph{allosteric gated pore alternating access mechanism} in which gate movement is strongly coupled to binding and the other gates.
Further examples will be shown in Section~\ref{sec:examples}.

\subsection{Gate states}
\label{sec:gatestates}

Thinking of transporters as consisting of $N$ gating elements that can individually switch between two states (such as open and closed as in Figure~\ref{fig:gatestates}) suggests a simple count to enumerate the possible number of conformations of the transporter,
\begin{gather}
  \label{eq:Nconf}
  n_{C} = 2^{N}.
\end{gather}
For a transporter with two gates, four states are possible, and eight states for $N=3$.\footnote{We equate a state with a conformation of the transporter, assuming that each state is formed by an distinguishable ensemble of conformers near the specific conformation. This typically implies that there is a kinetic separation between states. Although this is not necessarily always the case in practice, we will nevertheless use state and conformation interchangeably to keep the discussion simple.} The simple count ignores the fact that gate movement must be coordinated in some fashion. The type of coupling will depend on the individual molecule and may even depend on the substrate \citep{Henderson:2019aa} but for canonical transport one might want to assume that a leak state with all gates open plays no important role and so $n_{C} = 2^{N} - 1$.\footnote{One could write $n_{C} = 2^{N}-F$ where $0 \le F < 2^{N}$ is the number of ``forbidden'' conformations if one knew through other means which conformations were not accessible.}

The primary advantage of such a simple enumeration is to provide a framework in which to place experimentally or computationally observed conformations. \citet{Forrest:2011ys} proposed a similar classification with eight states, consisting of different conformations and with differing substrate occupancy. Their scheme makes use of thin gates but places central importance on the major conformational switch between inward and outward facing conformations. It has been successfully used to, for instance, categorize the wealth of structural data for the APC transporter BetP, for which crystal structures have been obtained for most of the states \citep{Ressl:2009la, Perez:2012sh, Perez:2014fk}, and to analyze simulated transitions for four LeuT-fold transporters \citep{Jeschke:2013lk}.

An almost trivial prediction of the gated pore model is the existence of \emph{occluded states}. In an occluded state, the transporter obtains a conformation in which the binding sites are not accessible from either compartment. In the doubly-gated pore, the occluded state naturally arises when the two gates are closed ($N=2$ in Figure~\ref{fig:gatestates}).
The alternating access model and the associated kinetic and thermodynamic analysis do not require occluded states for energy transduction and vectorial transport. Therefore, the existence of occluded states, which are not strictly necessary for function, could be interpreted as a consequence of the structural constraints of the implementation of alternating access (via inverted repeat symmetry) in proteins. Below we will show some structural evidence for occluded states. But it is also noteworthy to point out that molecular dynamics (MD) computer simulations have been able to generate occluded states when started from crystallographic conformations corresponding to inward or outward facing states: For example, \citet{Latorraca:2017aa} simulated the LbSemiSWEET transporter with unbiased MD and observed full transitions from outward to inward facing states that passed through an occluded state. The simulation spontaneously reached experimentally determined structures for inward open and occluded LbSemiSWEET. They found that the transitions were driven by favorable inter-helical interactions when either the extracellular or the intracellular gate closed and by an unfavorable helix configuration when both gates were closed. The two gates became tightly coupled and so prevented simultaneous gate opening, which would result in a leak state.
Other simulation examples are discussed below, which all point to the insight that molecular gates are a simple way to generate alternating-access states. In the absence of specific coupling that prevents two gates from closing at the same time, occluded states will occur.

\subsection{Examples of transporters as gated pores}
\label{sec:examples}

We will illustrate the physical reality of gates in transporters in three examples, corresponding to the common classification of the alternating access conformational transition as a rocker switch, as rocking bundle, and as an elevator movement \citep{Drew:2016aa}.

\subsubsection{Rocker switch transporters: MFS transporters: Two gates}
\label{sec:mfs}

\begin{figure*}
  \centering
  \includegraphics[width=\textwidth]{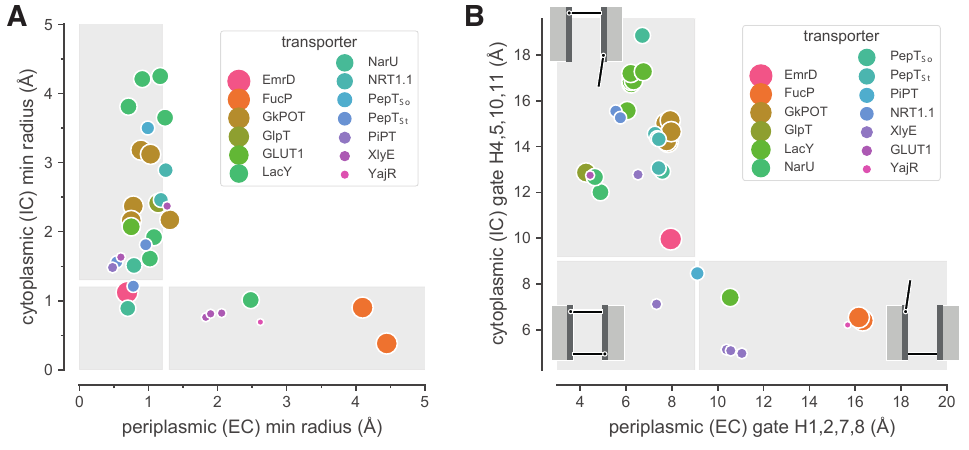}
  \caption{Conformational state of MFS transporters from a survey of crystal structures. The gray rectangles approximately indicate classification of the structures as outward open, occluded, or inward open.
    \textbf{A} Minimal pore radii $R$ computed with HOLE \citep{Smart96} near the periplasmic and cytoplasmic entrance. A conformation is classified as occluded if both $R \le 1.2$~\AA.
    \textbf{B} Gate distances $d$, calculated as the minimum distance between the C$_{\alpha}$ atoms of the relevant pairs of helix tips H1,2,7,8 (periplasmic gate) or H4,5,10,11 (cytoplasmic gate) \citep{Fowler:2015ix}. A conformation is classified as occluded for both $d \le 9$~\AA.
    Figure drawn after \citet{Fowler:2015ix}.
  }
  \label{fig:mfsconformations}
\end{figure*}

The MFS transporters all share a common fold with four inverted repeats (Figure~\ref{fig:invertedrepeats}A).
Based on inward facing and outward facing crystal structures, it was originally believed that alternating access would proceed by a rigid body movement whereby the two halves of the protein would move relative to each other in a rigid ``rocker switch'' manner \citep{Law:2008ny}. The discussion concentrated on LacY for which an alternative model described the protein as more flexible, with cytoplasmic and periplasmic openings governing access to the binding site, effectively describing gates \citep{Kaback:2007aa, Kaback:2019aa}. 
The existence of an occluded state in LacY would corroborate the gated pore model for its mechanism.
\citet{Stelzl:2014fk} hypothesized that LacY functioned as a pore with two coupled gates that could both close at the same time. With this assumption they could perform biased MD simulations to generate a model of occluded LacY with both gates closed. The model broadly agreed with experimental electron-electron resonance (DEER) spectroscopy data. More recently, experimental evidence for an occluded apo intermediate of LacY (based on sugar accessibility of a cysteine cross-linked mutant) \citep{Smirnova:2018aa} corroborated the occluded model.

The MFS transporter PepT\textsubscript{So} is a proton-coupled bacterial symporter for which only inward facing crystal structures are known \citep{Newstead:2011fk, Fowler:2015ix}. An open question has been the nature and molecular mechanism of the conformational transition between inward and outward conformation.
\citet{Fowler:2015ix} used an array of experimental and computational techniques, including X-ray crystallography,  DEER, and MD, to elucidate the dynamics of PepT\textsubscript{So}. They found that PepT\textsubscript{So} is representative for a large number of MFS transporters in that the ends of the first two helices in each of the four inverted repeats (see Figure~\ref{fig:invertedrepeats}A) form gates at the EC and IC entrance.
A wealth of structural data exists for the MFS transporters \citep{Drew:2021aa} with many transporter structures having been solved to atomic resolution in different conformations. \citet{Fowler:2015ix} analyzed 33 MFS structures in terms of the minimal pore radius near the periplasmic (extracellular, EC) and cytoplasmic (intracellular, IC) entrance (Figure~\ref{fig:mfsconformations}). The structure could neatly be categorized as outward facing, inward facing, or occluded, based either on geometrical constriction radius (Figure~\ref{fig:mfsconformations}A) or on the distances of the inverted-repeat gates (Figure~\ref{fig:mfsconformations}B).

Using multiple MD simulation with a Markov state model, \citet{Selvam:2018aa} sampled conformational transitions from the inward facing conformation through an occluded state to an outward facing conformation of PepT\textsubscript{So}. Their computational results were validated by comparison of simulated with experimental DEER spectroscopy data. The occluded state formed by closure of both ends of the protein. In a computed free energy landscape, the occluded conformation occupied a stable local minimum, as expected for a thermodynamic state.

The existence of occluded states (in crystal structures and MD simulations) in addition to the pseudo-symmetrical rocker switch conformations indicate that MFS transporters do not follow a strict rigid-domain rocker switch mechanism \citep{Drew:2021aa} but instead transition between these states in a more flexible manner (Figure~\ref{fig:realcyles}B).
The central binding site remains near the center of the transporter during the transport cycle while extracellular and intracellular gates change conformations locally in such a way that the end state conformations are approximately symmetry related.
Overall, the evidence suggests that MFS transporters can be described as transporters with two gates that are related by the internal repeat symmetry.

\subsubsection{Rocking bundle transporter Mhp1: Three gates}
\label{sec:mhp1triplegated}

\begin{figure}
  \includegraphics[width=\columnwidth]{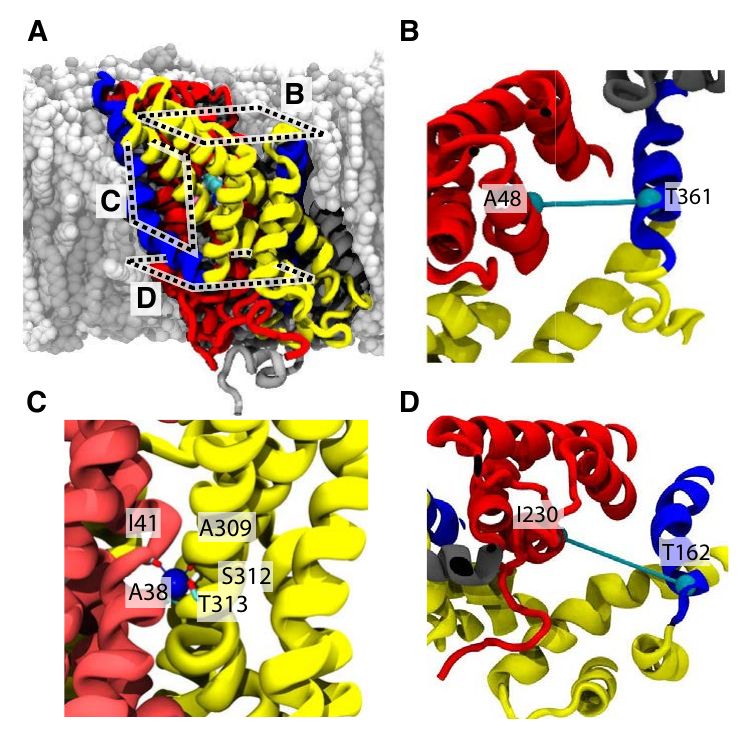}
  \caption{The three gates in Mhp1. \textbf{A} Mhp1 in the membrane.
    The hash motif (helices 3, 4, 8, 9) is shown in yellow, the bundle (helices 1, 2, 6, 7) in red, flexible (thin gate) helices 5 and 10 in blue, and C-terminal helices 11 and 12 in gray.
    The views on the gates (B--D) are indicated by broken rectangles.
    \textbf{B} extracellular thin gate (formed by TM10).
    \textbf{C} thick gate, quantified by the distance across the Na2 sodium binding site.
    \textbf{D} intracellular thin gate (TM5).
    Molecular images were produced with VMD \citep{Hum96}.
  }
  \label{fig:mhp1gates}
\end{figure}

\begin{figure*}
  \centering
  \includegraphics[width=\textwidth]{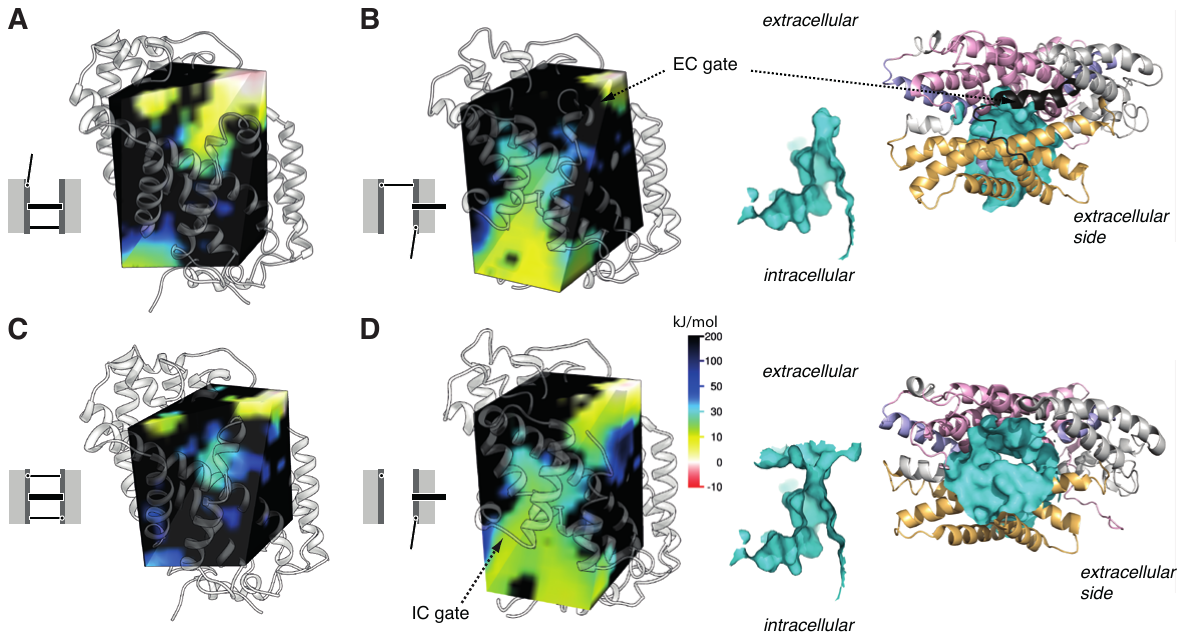}
  \caption{The putative role of the extracellular (EC) gate of Mhp1 is to prevent a sodium leak when the thick gate is open.
    The left four panels show a cut through the electrostatic solvation free energy (Born energy) landscape of a Na$^+$ ion inside Mhp1 for different conformations of the transporter, computed with the Poisson-Boltzmann equation. Red ($-10$~kJ$\cdot$mol$^{-1}$) to yellow ($+10$~kJ$\cdot$mol$^{-1}$) regions can be considered accessible for sodium ions under typical conditions.
    (The Na$^{+}$ Born energy was calculated as described previously \citep{Stelzl:2014fk}.)
    The gate cartoons in \textbf{A}--\textbf{C} represent some of the states for the triple-gated transporter in Figure~\protect\ref{fig:gatestates}; the cartoon in \textbf{D} symbolizes a leaky state that was artificially modeled by  removal of the EC-gate portion of TM10.
    \textbf{A} Outward facing open conformation (EC gate open, thick gate closed).
    \textbf{B} Inward facing open conformation (EC gate closed, thick gate and IC gate open).
    The solvent accessible surface (cyan) is shown from the side in the context of the protein helices (view on the surface from the top). The color scheme for the helices is the same as in Figure~\protect\ref{fig:mhp1gates}, except that the N-terminal half of the EC gate (TM10) is shown in black, and other colors are muted. The closed EC gate prevents a continuous sodium pathway.
    \textbf{C} Outward facing occluded conformation (EC gate closed, thick gate closed).
    \textbf{D} Simple model for a hypothetical inward facing open, leaky conformation with IC gate and thick gate open and EC gate removed. The solvent accessible surface representation and the electrostatic free energy show a sodium pathway through the membrane-spanning portion of the transporter.
    Molecular images were produced with UCSF Chimera \citep{Goddard:2007ys} and PyMOL \citep{PyMOL}.
  }
  \label{fig:surface}
\end{figure*}

\begin{figure*}
  \centering
  \includegraphics[width=\textwidth]{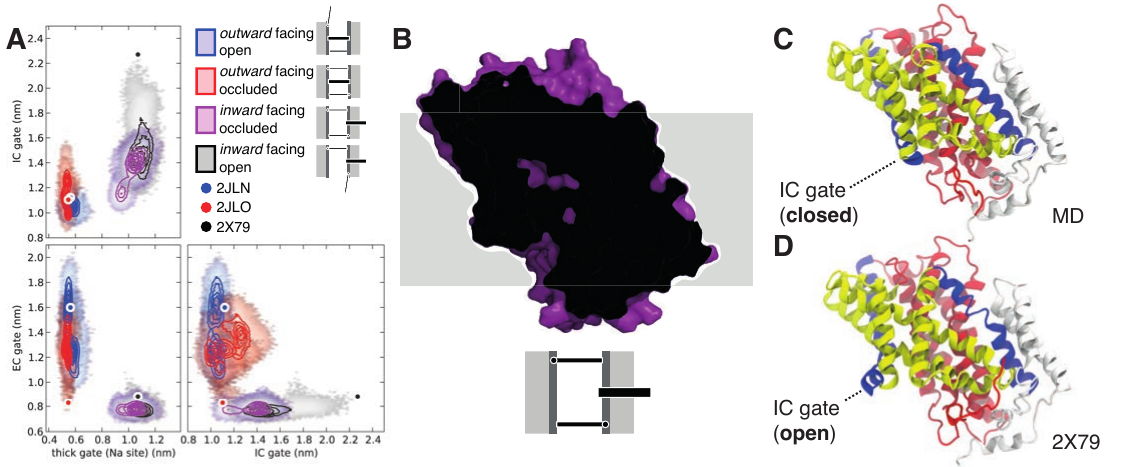}
  \caption{Molecular dynamics simulations sample an inward facing occluded state of Mhp1.
    \textbf{A} Gate distances (see Figure~\protect\ref{fig:mhp1gates}) from 100-ns MD trajectories of Mhp1 with a Na$^{+}$ ion in the Na2 site (simulations from \citet{Shimamura:2010uq} and additional unpublished data). Contour lines are drawn at 20\% increments of probability density. The shaded area indicates the full extent of order parameters explored in the simulation. Simulations started from crystal structures (circles) in three different conformational states (represented by the PDB ID), except for the magenta trajectories, which were started from a frame of the \emph{inward facing open} simulation (gray) that showed an almost closed intracellular gate. Data for three independent simulations each starting from the outward facing states (blue and red) are shown together with one from the inward facing open (gray) and two from the inward facing occluded (magenta) states.
    \textbf{B, C} Putative inward-facing occluded conformation of Mhp1 (snapshot from the high-probability region of the \emph{inward facing occluded} trajectory (magenta) in A). The cut through the transporter is shown in the same way as the crystal structures in Figure~\protect\ref{fig:mhp1states}.
    \textbf{D} Inward facing open crystal structure (PDB ID 2X79) with the open IC gate.
    Molecular images were drawn with PyMOL \citep{PyMOL} and VMD \citep{Hum96}.
  }
  \label{fig:mhp1iOCC}
\end{figure*}

In the rocking bundle transport mechanism, a transporter with two asymmetric halves undergoes alternating access by the movement of one domain relative to the other whereby the centrally located substrate binding site remains fixed \citep{Drew:2016aa}. The transporters in the large LeuT-like superfamily of APC (amino acid-polyamine-organoCation) transporters \citep{Vastermark:2014kl} were the first for which this mechanism was confirmed \citep{Kazmier:2017aa} and they have been characterized as
transporters with three gates \citep{Krishnamurthy:2009ij, Abramson:2009ci} (Figure~\ref{fig:realcyles}A).

Here we use the LeuT-like hydantoin permease Mhp1 from \emph{Microbacterium liquefaciens} as an example. Mhp1 is a nucleobase-sodium symporter (NCS1 family), which co-transports one sodium ion with one 5-substituted hydantoin \citep{Suzuki:2006fd}. It shares a five-helix inverted repeat architecture with other members of the superfamily of LeuT-like transporters \citep{Cameron:2013kx}.
X-ray crystallographic structures of Mhp1 in outward facing and inward facing conformations together with computer simulations, mass spectrometry, and electron paramagnetic resonance (EPR) measurements revealed the structural basis for the alternating access mechanism in this secondary transporter \citep{Weyand:2008gd, Shimamura:2010uq, Zhao:2013vn, Simmons:2014pt, Kazmier:2014uq, Calabrese:2017aa, Li:2019aa}.

The transporter can be understood as a gated pore with two thin and one thick gate \citep{Krishnamurthy:2009ij}, i.e., as a pore with three gates: The thick gate regulates the passage through the center of the membrane by means of the large conformational change---the rocking bundle motion---that switches the transporter from its outward facing to its inward facing conformation. In Mhp1 it consists of the hash motif (formed by helices TM3, TM4 and their inverted-repeat counterparts TM8 and TM9; see Figures~\ref{fig:mhp1gates}A, C and \ref{fig:invertedrepeats}B) that can rotate by about $30^{\circ}$ on an axis parallel to TM3 relative to the four-helix ``bundle'' (TM1, TM2 and TM6, TM7) \citep{Shimamura:2010uq}.
Thin gates are formed by the N-termini of the pseudo-symmetry related helices TM5 and TM10 and the linker to each preceding helix \citep{Shimamura:2010uq}. The extracellular (EC) thin gate (TM10; Figure~\ref{fig:mhp1gates}B) governs access to the substrate binding site from the periplasmic medium while the intracellular (IC; Figure~\ref{fig:mhp1gates}D) gate fulfills the symmetrical role of controlling the pathway to the cytosol.

The sodium binding site is formed between bundle and hash motifs, so opening of the thick gate, i.e., the alternating-access transition, opens up the sodium binding site and weakens ion binding to ensure rapid diffusion of the ion into the cytosolic compartment and opening a pathway for the substrate to follow \citep{Shimamura:2010uq, Zhao:2013vn}.
(In LeuT, the bundle rotates relative to the scaffold domain (equivalent to the hash motif in Mhp1) and helix TM1a undergoes a large outward bending motion \citep{Navratna:2019aa} as indicated in Figure~\ref{fig:realcyles}A, reinforcing the \textit{caveat} of structural biology that molecular mechanisms can depend sensitively on the actual molecular details.)

The role of the thin gates in Mhp1 appears to be more subtle (and almost certainly differs from the role of thin gates in related neurotransmitter sodium symporter (NSS)-like transporters such as LeuT \citep{Kazmier:2014uq, Kazmier:2014oq, Kazmier:2017aa}). 
In Mhp1, the whole N-terminus of TM10 moves together with the linker between TM9 and TM10 \citep{Weyand:2008gd, Shimamura:2010uq, Kazmier:2014uq}, thus forming a distinct gate structure that is mirrored in TM5 and the TM 4-5 linker, which are related to TM9/10 through the inverted-repeat symmetry as described in Section~\ref{sec:symmetry} and shown in Figure~\ref{fig:invertedrepeats}B. 

As discussed in Section~\ref{sec:alternatingaccess}, a protein that functions according to the alternating access mechanism cannot function if it presents a continuous, leaky pathway \citep{Tanford:1983ek} [or if it allows too many nonproductive leak cycles to occur (Section~\ref{sec:thermodynamics})].
The EC gate appears to prevent Mhp1 from leaking the driving ion, Na$^{+}$, as demonstrated by modeling:
Figure~\ref{fig:surface}B shows that in the inward facing conformation, with the EC gate closed, the solvent accessible surface only extends from the IC side into the binding site at the center of the protein. However, when the EC gate is removed (the atoms were deleted from the structure as a simple model of a hypothetical inward facing conformation with an open EC gate) a pathway opens up through the membrane (Figure~\ref{fig:surface}D). The calculated electrostatic solvation free energy (Born energy) in the volume of the pathway shows that Na$^{+}$ ions could traverse the membrane because a low-energy path is visible (Figure~\ref{fig:surface}D). On the contrary, in all other states, no low energy path can be found for a sodium ion (Figures~\ref{fig:surface}A--C) because either the thick gate or the EC gate blocks the passage. Thus, the EC gate fulfills an important role in preventing a sodium ion leak when the thick gate is in its open, i.e., inward facing state. When the thick gate is closed (outward facing), the EC gate only regulates access to the binding site from the extracellular compartment.

Additionally, the EC gate is involved in substrate selectivity \citep{Simmons:2014pt}. Mhp1 transports 5-substituted hydantoins where the substituent must be a bulky hydrophobic moiety such as a benzyl or indolylmethyl group. However, if the 5-substituent is too voluminous such as a naphtyl group, transport is inhibited even though the molecule binds tightly. A crystal structure of outward-facing Mhp1 with 5-(2-naphthylmethyl)-L-hydantoin bound revealed that the EC gate was trapped in an open conformation due to a steric clash of the naphtyl ring with Leu363 \citep{Simmons:2014pt}. A Leu363Ala mutant of Mhp1, which removes the clash, was competitive for transporting 5-(2-naphthylmethyl)-\textit{L}-hydantoin. These results strongly suggest that closure of the EC gate is required for the alternating access transition to occur. In the language of the gated pore view, the thin EC gate is coupled to the thick gate. Structural comparison suggests that this coupling is due to the geometrical architecture and the direct connection of the rotating hash motif to the EC gate through the 9-10 linker. The thick gate cannot move into the space occupied by the open EC gate and therefore is prevented from closing. Conversely, the open thick gate appears to latch the EC gate in its closed position \citep{Shimamura:2010uq}, possibly helped by changes in the extracellular EL4 loop \citep{Song:2015aa}. In LeuT, a different volume sensing mechanism was described, using a combination of MD simulations and single molecule FRET experiments, whereby two hydrophobic residues, F259 and I359, allosterically couple occupancy of the binding site with a substrate of the right size to release of a sodium ion and subsequent conformational transition to the inward-facing state \citep{LeVine:2019aa}.

The presence of three gates naively predicts the existence of $2^{3}-1 = 7$ distinct conformations (discounting variations due to bound substrate and ions and the leak state). In the following we will discuss in how far Mhp1 follows this prediction and how additional considerations about coupling between gates reduce the number of distinct states.

The crystallographic structures of Mhp1 in Figure~\ref{fig:mhp1states} show an occluded state \citep{Weyand:2008gd, Simmons:2014pt} in addition to the outward \citep{Weyand:2008gd} and inward facing states \citep{Shimamura:2010uq} that are necessary for alternating access. Given the definition of the three gates (see Figure~\ref{fig:mhp1gates}), the crystallographic occluded structure is in an outward-facing occluded conformation because the thick gate is closed (outward facing) and both thin gates are also closed. MD simulations had shown that the thin gates could change conformations on the 100-ns timescale \citep{Shimamura:2010uq}. A detailed analysis of the simulations in terms of the gate distances (Figure~\ref{fig:mhp1iOCC}A) showed that the EC gate was mobile when the thick gate was closed (outward facing conformation) and the simulations sampled both outward open and outward occluded conformations. The IC thin gate remained locked, though. Conversely, once the thick gate was open (inward facing conformation), the EC gate was locked and the IC gate could sample open and closed conformations. The dynamic behavior of the thin gates reflects the two-fold symmetry that is imposed by the inverted repeat symmetry (Figure~\ref{fig:invertedrepeats}B).
Using site-directed spin labeling and DEER spectroscopy, \citet{Kazmier:2014uq} showed that the IC gate formed by TM5 undergoes motions between open and closed conformations, in agreement with the computational results.

The crystallographic inward facing structure\citep{Shimamura:2010uq} (PDB ID 2X79) shows the IC gate in a wide open position\footnote{The crystals that produced the inward facing open structure with PDB ID 2X79 only formed when cells were grown on minimal medium with seleno-\textit{L}-methionine \citep{Shimamura:2010uq}. The electron density map of 2X79 contains a blob of unidentified density in the inward facing cavity that appears to have wedged open the IC gate and stabilized the IF conformation. It is possible that the unidentified molecule(s) held the IC gate in an especially wide open position from which it relaxes in the MD.}, corresponding to an inward open structure. The MD simulations with the thick and both thin gates closed showed that a second occluded state might exist (Figure~\ref{fig:mhp1iOCC}B,C), as predicted by a pore with three gates (Figure~\ref{fig:gatestates}).
Thus, the MD simulations sampled one additional conformation predicted for a triple-gated pore, bringing the total observed to 4 out of 7 (discounting the leak state). They also suggest that some other conformations are not observable because of coupling between thick and thin gates. For example, opening of the thick gate while the EC gate was open was experimentally ruled out \citep{Simmons:2014pt} (see also the preceding discussion of volume sensing); conversely the MD suggested that the EC gate will remain closed when the thick gate is open. These observations rule out two more putative states (EC open/thick open/IC closed and the leak state, EC open/thick open/IC open). The remaining two states (EC closed/thick closed/IC open and EC open/thick closed/IC open) seem unlikely based on the MD, which showed that the IC gate is locked when the thick gate is closed, so that a closed thick gate prevents an open IF gate. Based on this analysis, only 4 out of 8 possible states, namely the three crystallographic conformations (OF open, OF occluded, IF open) and the MD-based prediction for IF occluded, should be the observable conformations for Mhp1. Further experiments and simulations will be necessary to confirm the coupling between the three Mhp1 gates as predicted from qualitative observations.

An analysis based on gate states is general but limited in important details and must be augmented with additional information about the relative stability of the states. For example, under physiological conditions, the inward facing conformation is more prevalent than the outward facing one although this can be changed with the addition of substrate \citep{Kazmier:2014uq, Calabrese:2017aa}. The DEER experiments suggest that some conformations such as the inward facing occluded one might be much shorter lived than other ones \citep{Kazmier:2014uq}. Such quantitative information is crucial in order to interpret kinetic reaction diagrams based on the predicted states, as discussed in Section \ref{sec:unified}.

\subsubsection{Elevator transporters: ASCT2 and CNT\textsubscript{NW}}
\label{sec:elevators}

\begin{figure*}
  \centering
  \includegraphics[width=\textwidth]{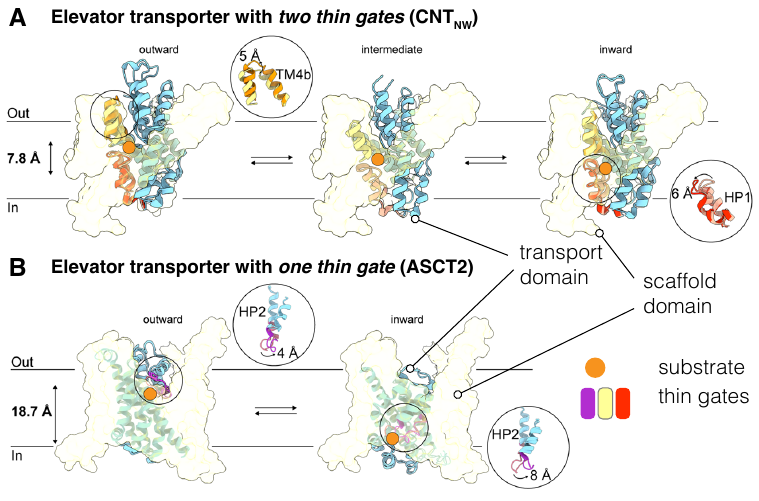}
  \caption{Elevator transporters with gates. The moving transport domain, which contains the substrate and ion binding sites (orange circle), is shown as colored helices. The static scaffold domain is shown in outline. Insets show structural elements that fulfill the role of thin gates. The large movement of the transport domain relative to the scaffold domain can be considered the thick gate, with the overall vertical movement indicated on the left. \textbf{A} The concentrative nucleoside transporter CNT\textsubscript{NW} of the SLC28 family uses two different gates, TM4b in the outward facing state [5-Å movement between closed (yellow) and open (orange)]  and HP1 in the inward facing one [6-Å movement between closed closed (light pink) and open (red)]. \textbf{B} In the neutral amino acid transporter ASCT2 of the SLC1 family, the \emph{same} helical hairpin 2 (HP2) controls access to the binding site in the outward facing [4-Å movement between closed (light pink) and open (magenta)] \emph{and} inward facing conformation [8 Å movement between closed (light pink) and open (magenta)].   (Figure modified with permission from A.\ A.\ Garaeva and D.\ J.\ Slotboom\protect\citep{Garaeva:2020aa}, Biochem.\ Soc.\ Trans., 48, 1227–1241 (2020). Copyright 2020 the Author(s); licensed under the Creative Commons Attribution 4.0 International (CC BY 4.0) license.)}
  \label{fig:elevator-gates}
\end{figure*}

A diverse class of transporters uses an \emph{elevator} mechanism to achieve alternating access \citep{Drew:2016aa, Garaeva:2020aa, Diallinas:2021aa}. A \emph{transport} domain moves relative to another static \emph{scaffold} domain up and down through the membrane and in this way exposes access to the binding site that resides fully (or mostly) in the transport domain \citep{Drew:2016aa}. Confirmed examples of elevator proteins include, for example, the bacterial homolog of a neuronal sodium/glutamate transporter Glt\textsubscript{Ph} \citep{Reyes:2009aa, Ruan:2017aa} (see Figure~\ref{fig:realcyles}C), the sodium/proton antiporter TtNapA \citep{Coincon:2016ly}, the divalent anion sodium symporter (DASS) VcINDY \citep{Mulligan:2016aa}, the concentrative nucleoside transporter (CNT) CNT\textsubscript{NW} \citep{Hirschi:2017aa},  the neutral amino acid transporter ASCT2 \citep{Garaeva:2019aa}, and the sodium/proton exchanger hsNHE1 \citep{Dong:2021aa}, with an extensive list of proposed elevator-like transport mechanisms provided by \citet{Garaeva:2020aa}. As for the rocking-bundle transporters, the large conformational elevator transition that switches between outward and inward facing conformation can be designated the thick gate, with the understanding that ``gate'' stands for a bistable element. Thin gates have been directly identified in a number of elevator transporters \citep{Garaeva:2020aa}. Two examples, CNT\textsubscript{NW} and ASCT2 are shown in Figure~\ref{fig:elevator-gates}; both are inverted-repeat trimeric membrane proteins whose protomers consist of a trimerization domain that forms the scaffold and a mobile transport domain. However, the oligomerization state does not appear to be a deciding element because gates have also been described for monomeric and dimeric elevator transporters \citep{Garaeva:2020aa}.

CNT\textsubscript{NW} is a sodium-driven nucleoside symporter for which multiple conformations---outward facing, multiple intermediates, and inward facing---were resolved with X-ray crystallography \citep{Hirschi:2017aa}. The structures showed that access to the binding site in the outward facing conformation is governed by a conformational change of transmembrane helix TM4b whereas in the inward facing conformation, helical hairpin HP1b fulfills the equivalent role (Figure~\ref{fig:elevator-gates}A). Therefore, with TM4b and HP1b considered as thin gates, and together with the elevator movement that switches between inward and outward facing conformations, CNT\textsubscript{NW} is an elevator transporter with three gates. Consistent with this categorization, \citet{Hirschi:2017aa} found inward-open, inward-occluded, outward open, and three slightly different intermediate occluded structures. Although the occluded structures have the two thin gates closed and so prevent an ion leak, they do not fit neatly into the simple scheme in Figure~\ref{fig:gatestates} with $N=3$ because the position of the transport domain is in an intermediate position between outward facing and inward facing instead of either as would be expected of a simple switch. The structures show that the transport domain undergoes internal conformational changes instead of a simple rigid body motion during the elevator transition \citep{Hirschi:2017aa}, somewhat reminiscent of the situation in the unrelated MFS transporters discussed above where N and C domain show internal motions. At least for elevator-type transitions the characterization of the transport domain movement as a ``thick gate'', i.e.{} as a simple bistable switch, may be an oversimplification that needs to be carefully considered. It nevertheless serves as a useful starting point to categorize conformations.

Elevator transporters also defy the categorization as simple gated pores in other ways. For example, in the Alanine Serine Cysteine Transporter 2 ASCT2, an exchanger for neutral amino acids \citep{Garaeva:2019aa}, the transport domain moves such a large distance through the membrane that the same hairpin HP2 can act as both the extracellular \emph{and} the intracellular thin gate (Figure~\ref{fig:elevator-gates}B), thus making ASCT2 a transporter with nominally one thin and one thick gate. Functionally, however, it is effectively a triply gated transporter because access to the binding site is controlled by HP2 in the inward facing and outward facing conformation \citep{Garaeva:2019aa, Garaeva:2020aa}. Consistent with this classification, at least four distinct conformations have already been determined experimentally, namely outward-facing open and occluded \citep{Yu:2019aa}, inward-facing occluded \citep{Garaeva:2018aa}, and inward facing open \citep{Garaeva:2019aa}. ASCT2 belongs to the SLC1 transporter family, the same as the neuronal glutamate EAAT transporters and their bacterial homolog Glt\textsubscript{Ph}, the archetypical elevator transporter \citep{Reyes:2009aa}. Recent structural evidence \citep{Wang:2020aa, Huysmans:2021aa} suggests that Glt\textsubscript{Ph} also employs HP2 as a dual-function gate in outward facing and inward facing conformations, similar to what was found for ASCT2.

\section{Unified transport cycle model}
\label{sec:unified}

\begin{figure}
  \centering
  \includegraphics[width=\columnwidth]{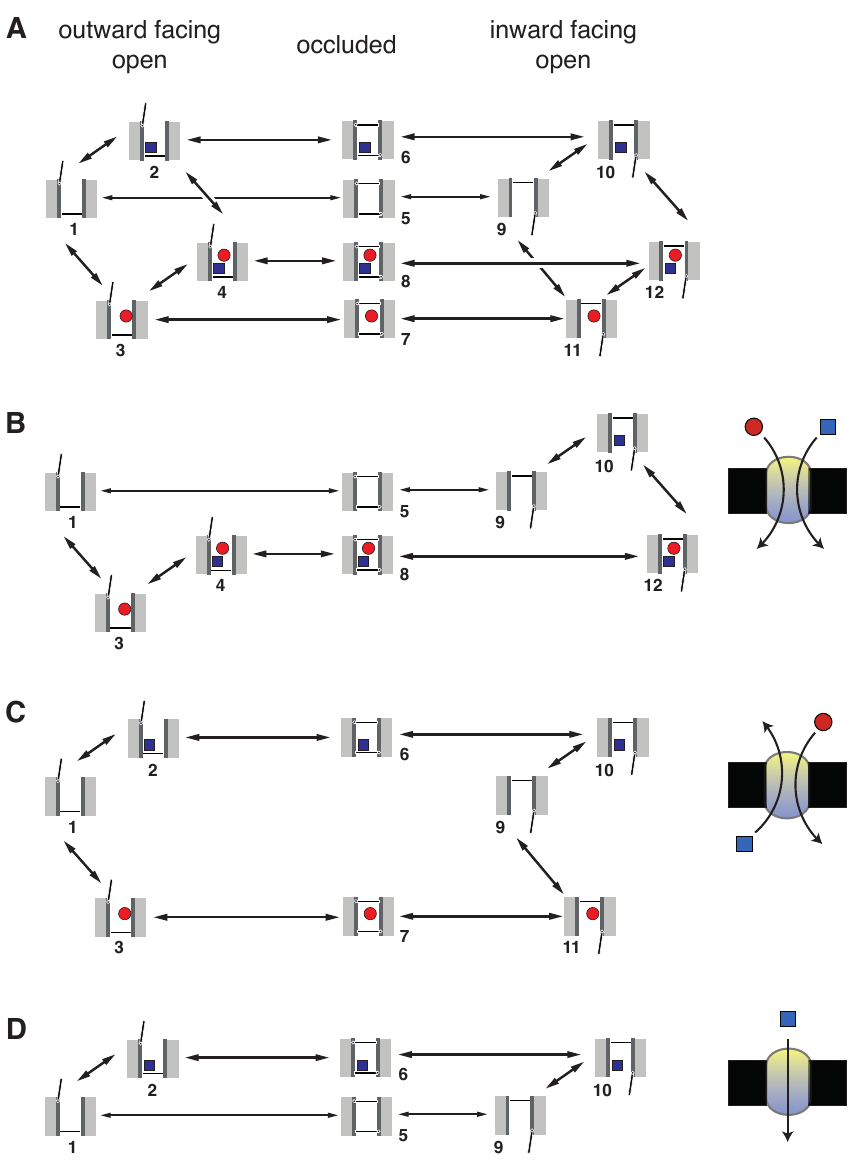}
  \caption{Unified picture of transporter function for a hypothetical transporter with a 1:1 stoichiometry between driving ion (red circle) and substrate (blue square) and two gates. All combinations of conformational states (outward facing open, inward facing open, and occluded) with bound ion and/or substrate are listed. Leak states are omitted for simplicity. When an ion or substrate is shown, the corresponding binding or dissociation reaction is implied. Depending on the physiological function (symbol on the right), only certain sequences of states are visited (in the idealized case) while others (omitted) are not part of the cycle.
    \textbf{A} Unified transport cycle with all possible kinetic paths.
    \textbf{B} Symporter. (The cycle drawn here corresponds to the one in Figure~\protect\ref{fig:transportcycle}A and a specific binding/unbinding order was assumed.)
    \textbf{C} Antiporter. (The cycle corresponds to the one in Figure~\protect\ref{fig:transportcycle}B with a specific binding/unbinding order.)
    \textbf{D} Uniporter.
  }
  \label{fig:unified}
\end{figure}

In our enumeration of gate states we have implicitly assumed that these states do not depend on the binding of driving ions or substrate. Such an assumption is not warranted. For instance, the symporter in Figure~\ref{fig:transportcycle}A must avoid leak cycles that involve a slippage transition between the inward- and outward-facing transporter conformations when when only the ion is bound or it would just dissipate the ion gradient. Mechanistically, the absence of the substrate when the ion is present must be changing the free energy landscape of the transition \citep{Lauger:1980qf} in such a way that the ion-only transition faces a much higher barrier than the fully loaded transporter. In other words, binding of an ion \emph{and} a substrate unlocks the transporter and enables the conformational transition to occur as experimentally found for the aspartate-sodium symporter Glt\textsubscript{Ph} \citep{Akyuz:2015il}.
A similar argument can be made for the antiporter cycle in Figure~\ref{fig:transportcycle}B where the transition between the empty inward and outward-facing states (Figure~\ref{fig:antiportercycle}A) needs to be suppressed to avoid leak cycles that only dissipate the ion. In this case, binding of the ion \emph{or} the substrate unlocks the transporter.
In both cases it is clear that presence or absence of bound ion and/or substrate corresponds to a protein with different energetics from, say, the unloaded conformation.

Consequently, one may roughly estimate the number of available states of a transporter as the product of the number of conformational states $n_{C}$ with the number of ways to bind driving ions and substrates. We can estimate $n_{C}$ from the gate model as $n_{C} = 2^{N}-1$ for $N$ gates where a leak state is excluded. As an example we assume an ideal transporter with a fixed ion:substrate stoichiometry of $\nu_{I}$:$\nu_{S}$. We further make the simplifying assumption that there are $\nu_{I}$ ion binding sites and $\nu_{S}$ substrate binding sites. Under these assumptions there are $n_{j} = \sum_{k=0}^{\nu_{j}}\binom{\nu_{j}}{k} = 2^{\nu_{j}}$ ways to distribute zero to $\nu_{j}$ particles over $\nu_{j}$ binding sites. Hence the number of states in this particular model is 
\begin{equation}
  \label{eq:nstates}
  \Omega(N, \nu_{I}, \nu_{S}) = n_{C}n_{I}n_{S} = (2^{N}-1)\,2^{\nu_{I} + \nu_{S}}.
\end{equation}
For example, for the transporters in Figure~\ref{fig:transportcycle}, $\nu_{I} = \nu_{S} = 1$ and with $N=2$, $\Omega = 3\times 4 = 12$ different states should be considered, as shown in Figure~\ref{fig:unified}.
In principle, all theoretically possible transitions between states (arrows in Figure~\ref{fig:unified}) contribute to an overall transport process \citep{Hill:1989ve}.
In our example, occluded states are included for completeness. However, occluded states (numbers 5--8) cannot exchange with each other and their effect could be replaced with an effective rate constant between the outward facing and inward facing conformations that are connected by the occluded state.

Equation~\ref{eq:nstates} illustrates how one can approach enumerating relevant states for a transporter by clearly stating the assumptions and then evaluating the resulting combinatorics. In realistic proteins these simplifying assumptions do not always hold. If ions and substrates compete for overlapping binding sites then $\Omega$ is less than the value given by Eq.~\ref{eq:nstates}. For instance, in the competitive binding mechanism observed in the sodium/proton antiporter NhaA \citep{Mager:2011nx, Calinescu:2015kl}, two protons $\nu_{I}=2$ and one sodium ion $\nu_{S}=1$ compete for two binding sites. Based on simulations \citep{Huang:2016aa}, one binding site can carry one of the protons ($n_{I_{1}} = 2^{1} = 2$) whereas for the second site, the other proton and the sodium ion compete ($n_{I_{2},S} = 2^{2}-1 = 3$), resulting in $n_{I_{1}} n_{I_{2},S} = 6$ potential states for each conformation. These binding site occupancies have been observed in simulations \citep{Lee:2014oq, Huang:2016aa} but it remains an open question how the mixed occupancy state (with one proton and one sodium ion) is prevented from transitioning between inward facing and outward facing conformation. Further complications arise when the stoichiometry itself varies depending on external conditions such as substrate identity or concentration. For example, the proton-coupled peptide transporters (POT) PepT\textsubscript{St} transports tri-peptides with a 3:1 proton:peptide stoichiometry whereas di-peptides are transported with 4 or 5 protons \citep{Parker:2014qf}. Another example is the mammalian sodium/iodide symporter (NIS) with a 2:1 Na$^{+}$:I$^{-}$ stoichiometry \citep{Eskandari:1997aa}. However, it transports perchlorate ClO$_{4}^{-}$ with the different 1:1 stoichiometry because perchlorate binds to an allosteric site that inhibits binding of one sodium ion to one of the two sodium binding sites; even micromolar concentrations of perchlorate alter the stoichiometry of iodide transport \citep{Llorente-Esteban:2020aa}. A second substrate binding site may also be present in the LeuT transporter that can positively stimulate intracellular gate opening \citep{Shi:2008dk}, thus necessitating more involved models as discussed by \citet{LeVine:2016aa}. The situation appears to be even more complicated for the Nramp transition metal transporters, which display different proton:metal stoichiometries depending on membrane voltage and chemical identity of the metal ion \citep{Bozzi:2019aa, Bozzi:2021aa}. In principle, these additional complexities can be taken into account in a kinetic model with the addition of additional states but they require detailed insights into the specific system in question.

The simple model for a transporter with two gates and ideal 1:1 stoichiometry generates, based on the state count, Equation~\ref{eq:nstates}, a kinetic scheme in which 12 states (four outward facing open, four occluded, and four inward facing open) are connected by either ion or substrate binding/unbinding reactions or conformational transitions (Figure~\ref{fig:unified}A).
This simple scheme already contains rich functional behavior because depending on the magnitudes of the \emph{rate constants} of different transitions, distinct physiological functions are realized \citep{PLoC:2019ab}.
If the conformational transitions with only substrate bound ($2 \leftrightarrow 6 \leftrightarrow 10$) and only ion bound ($3 \leftrightarrow 7 \leftrightarrow 11$) are suppressed by virtue of low kinetic rate constants \citep{Hill:1981aa}, then naturally the transport cycle for a symporter (Figure~\ref{fig:unified}B) emerges.
Conversely, if the apo transition ($1 \leftrightarrow 5 \leftrightarrow 9$) and the transition with both substrate and ion bound ($4 \leftrightarrow 8 \leftrightarrow 12$) are suppressed, the cycle represents an antiporter (Figure~\ref{fig:unified}C).
The same observation had been made by \citet{Robinson:2017aa}, who postulated such a \emph{free exchange model} for the EmrE transporter.
Furthermore, by suppressing transitions that involve the driving ion ($3 \leftrightarrow 7 \leftrightarrow 11$ and $4 \leftrightarrow 8 \leftrightarrow 12$) and only retaining a cycle that contains the alternating-access transition with either substrate bound or the empty transporter, a simple uniporter model emerges (Figure~\ref{fig:unified}D). In this case, no energy coupling occurs and the substrate will move down its electrochemical gradient by facilitated diffusion, as observed, for instance, in some sugar transporters \citep{Henderson:2019aa, Drew:2021aa}.

The cycles in Figure~\ref{fig:unified}B--D were purposefully simplified and only highlighted one specific sequence of substrate and ion binding and unbinding events. It is not always clear that the binding order is fixed. The model can be made more realistic by retaining all transitions related to ion and substrate binding (connections between outward facing states 1, 2, 3, and 4 and between inward facing states 9, 10, 11, and 12 in Figure~\ref{fig:unified}) while still suppressing the undesirable conformational transition (e.g., the ion leak pathway $3 \leftrightarrow 7 \leftrightarrow 11$ for the symporter in Figure~\ref{fig:unified}B or the apo transition $1 \leftrightarrow 5 \leftrightarrow 9$ for the antiporter in Figure~\ref{fig:unified}C). With these additional transitions included, different sequences of binding or dissociation reactions would all be considered simultaneously and, depending on ion and substrate concentrations, different binding orders may prevail \citep{Hussey:2020aa}. Overall, the three physiologically distinct types of transporters primarily differ in which of the alternating-access transitions are forbidden (or strongly suppressed). Mutations may differentially change the transition rates and so switch the function of transporter from, say, a symporter to a uniporter as seen in the MFS sugar transporters \citep{Madej:2014fc} or Nramp transporters \citep{Bozzi:2021aa} or turn an antiporter into a symporter as reviewed by \citet{Henderson:2019aa}.
Furthermore, different transport modalities may depend on the chemical identity of the substrate. For example in Nramp transporters, the type of metal ion controls whether either cotransport with protons occurs (with a stoichiometry that may vary depending on voltage and pH) or metal ion uniport \citep{Bozzi:2021aa}. Although the kinetic scheme shown in Figure~\ref{fig:unified}A does not model different stoichiometries, it is not difficult to extend it with additional states and paths for, say, multiple ion binding/unbinding reactions in order to assess transport as a function of external parameters such as pH, substrate concentrations, and membrane potential.

The presence of occluded states in the transport cycle is important because the alternating access conformational transition cannot complete if the occluded intermediate state cannot be reached. As the examples of volumetric sensors in Mhp1 \citep{Simmons:2014pt} and LeuT \citep{LeVine:2019aa} show (see Section~\ref{sec:mhp1triplegated}), some proteins control the conformational switch by kinetically controlling progression through an occluded state. As another example, the transport domain of the CNT\textsubscript{NW} transporter contracts in intermediate states in ordered to facilitate the apo transition of its symporter cycle \citep{Hirschi:2017aa}. Overall, the simple model indicates that occluded states could act as convenient control points for \emph{kinetic control} of function (control of rates as performed by enzymes) as opposed to \emph{thermodynamic control} (changes in binding affinities and free energy differences between states). 

The unified view of transporter function has recently been investigated in more detail.
The Henzler-Wildman group ran computational models that indicated that, given the known parameters of the EmrE transporter \citep{Robinson:2017aa}, the protein should be able to switch between symport and antiport mode based on external concentrations \citep{Hussey:2020aa}.
In a commentary on the EmrE work, \citet{Grabe:2019aa} outlined how slippage processes should be considered an important part of transporter function.
They then went on to demonstrate computationally how slippage pathways can provide a form of proof-reading and a new kind of selectivity mechanism for transporters \citep{Bisignano:2020aa}.
Specifically, they showed how the bacterial vSGLT sodium:sugar symporter might be able to discriminate carbohydrate substrates from very similar toxic compounds and under certain conditions even export toxins.
Furthermore, Zuckerman and co-workers employed a systematic approach to explore the space of all transport reactions (such as Figure \ref{fig:unified}A for the simple 2-gate, 1:1 transporter) to computationally engineer transporters with specific functions such as antiport or symport with specific selectivities \citep{George:2020aa}.
Overall, there is increasing evidence that the unified transport cycle model is a useful and comprehensive framework in which to understand the continuum of secondary transporter function.

\section{Towards biological complexity}
\label{sec:limitations}

In order to be general and to delineate overall constraints under which transporters operate and to show some common patterns, the preceding discussion omitted many details that are nevertheless essential for actual transporter proteins. There are only a few hard rules that biological secondary active transporters have to adhere to: The only available free energy sources are external electrochemical gradients and vectorial transport as an out-of-equilibrium process is governed 
by the second law of thermodynamics, as outlined in Section~\ref{sec:thermodynamics}. However, thermodynamics does not say anything about molecular details and even leaves some loopholes open in the form of the possibility for ergodicity-breaking processes that have been hypothesized to be at work in electron-transport proteins \citep{Matyushov:2015vn}. Some of the exceptions and deviations from the general principles have already been mentioned in passing. In order to provide a compact summary and to point towards the considerations required to move towards experimentally observed biological complexity, a number of findings will be reiterated.

Not all secondary active transporters fall neatly in the scheme described above. Inverted repeat symmetry is often observed but this is not a necessary requirement because bi-stable proteins can conceivably evolve by other routes, too. Some transporters (such as chloride/proton antiporters, ClC) do not undergo distinct large scale (domain-level) conformational changes and instead access to the binding site is governed at the single side chain level \citep{Feng:2010sw, Basilio:2014aa, Mayes:2018aa}.

The key questions in understanding the molecular mechanism of transport are (1) where do ions and substrate bind and what are energetics and kinetics of the binding/dissociation processes, (2) what are the moving elements of the alternating access conformational transition and what are the rates, and (3) how are the interactions of the driving ions and transported substrates coupled to the conformational transition \citep{Rudnick:2013aa, LeVine:2016aa}? Counting states and connecting them via binding or dissociation reactions or conformational transitions provides a framework within which to organize experimental and computational findings to answer (1) and (2). However, the mechanism of coupling between conformational change and ion/substrate will always require a detailed analysis of the specific transporter, as shown, for example, for LeuT-like transporters \citep{Stolzenberg:2015aa, Zhang:2018ae}, ClC \citep{Mayes:2018aa}, or as outlined for LacY \citep{Kaback:2015vn}.

In principle, a kinetic diagram \citep{Hill:1989ve} or a master equation approach \citep{Zwanzig:2001aa} can capture the non-equilibrium dynamics of an arbitrary complex process, as long as it is possible to define the different states and obtain the appropriate rates. For instance, a kinetic diagram with the appropriate states included can yield very different forward and backward transport rates \citep{Zhang:2007ab} and exhibit switching between symporter and antiporter functionality \citep{Hussey:2020aa} and kinetic selectivity \citep{Bisignano:2020aa}. Such diagrams will contain different conformational change rates and different on/off rates, which in turn requires different molecular environments in OF and IF conformations. The kinetic diagram approach is also flexible enough to model transporters where co- and countertransport occur together, as for instance for some bacterial homologs of NSS transporters (MhsT, Tyt1) that co-transport Na$^{+}$ and substrate and counter-transport H$^{+}$ \citep{Zhao:2010aa, Malinauskaite:2016aa, Fan:2021ab}, the serotonin transporter SERT, which co-transports Na$^{+}$, serotonin, and Cl$^{-}$ and counter-transports K$^{+}$ or H$^{+}$ \citep{Keyes:1982aa}, or for the neuronal glutamate transporter EAAT3, in which inward-directed Na$^{+}$ and H$^{+}$ gradients and the outward facing K$^{+}$ gradient drive glutamate uptake \citep{Zerangue:1996aa}.  Similarly, by adding additional states to the diagram it would also be possible to model transporters with stoichiometries that differ with the substrate, such as the POT \citep{Parker:2014qf} and Nramp \citep{Bozzi:2021aa} transporters. Expanding diagrammatic cycle models to accommodate increasing complexity and realism comes at the cost of an increasing number of states and rates that need to be measured or estimated, as seen for the multiscale kinetic model of ClC-ec1 with $N=2^{6}=64$ states \citep{Mayes:2018aa}. However, even though \textit{a priori} $N(N-1) = 4032$ rate coefficients would be needed for this model, only 68 rate coefficients had to be actually determined because most transitions were disallowed as the states were chosen to represent elementary processes such as single binding/dissociation events. Thus, with a good choice of states, kinetic diagrams appear to be a viable approach to represent even complex transport cycles.

The next step in modeling transporter function should take into account the interaction of the membrane protein with its environment. These interactions may include the effect of the lipid bilayer on the transporter \citep{Denning:2013ly, Laganowsky:2014vn, Koshy:2015kx, Ernst:2021aa}. They may also include the effect of regulation through phosphorylation \citep{Ramamoorthy:2011kl, Foster:2017ab, Parker:2014zr}, which may take the form of phosphorylated states having very different transport properties than the unphosphorylated ones, as shown for the human dopamine transporter hDAT where interaction of the phosphorylated  N-terminus with the transport pathway may affect Na$^{+}$ efflux \citep{Moritz:2015aa, Razavi:2018aa}. Finally, realistic models should account for the effect of allosteric modulation by small molecules such as occupancy of the allosteric site in neuronal neurotransmitter transporters \citep{Navratna:2019aa} or the iodide transporter \citep{Llorente-Esteban:2020aa}, or potentiators in neuronal glutamate transporters \citep{Rives:2017aa, Kortagere:2018aa}.

\section{Conclusion}
\label{sec:conclusion}

We provided a perspective on broad and general principles that apply to many secondary active transporters.
Transporters are seen as catalysts or ``physical enzymes'' that enable transport across the cell membrane against an electrochemical gradient by transducing free energy from the electrochemical gradient of a driving ion to the vectorial transport of the substrate. Energy transduction requires cyclical reactions that include both driving ion and transported substrate. The alternating access model provides a simple scheme through which such cycles can be established. The protein must exist in at least two distinct conformations in which the binding sites are exposed only to either the outside or the inside solution. Importantly, no energy transduction is possible if a continuous pore is established. The two protein conformations that are needed for the two alternating access states are often related by a structural two-fold pseudo symmetry that originates in inverted repeats in the protein's genetic sequence. Currently, three broad classes of alternating access transitions have been described: rocker-switch, rocking-bundle, and elevator transitions. A description of transporters as gated pores is fruitful in many cases because gates (two state switches) can be identified with structural elements in the transporter. Enumerating all distinct gate states naturally includes occluded states in the alternating access picture and also suggests what kind of protein conformations might be observable. By connecting the possible conformational states and ion/substrate bound states in a kinetic model, a unified picture emerges in which symporter, antiporter, and uniporter function are extremes in a continuum of functionality.
Although not all transporters can be as neatly classified as some of the examples given, the kinetic diagram approach provides a flexible and quantitative framework in which to organize our knowledge from experiments and simulations.

Many open questions remain. For example, the molecular mechanism of coupling between conformational changes and ion/substrate binding and allosteric interactions between gates need to be evaluated for most known transporters. General theories of allosteric coupling will likely be helpful to define the specific quantitative questions that need to be asked \citep{LeVine:2016aa}. Occluded states were explained as a consequence of the existence of gates in transporters, so a natural question to ask is if all transporters have occluded states, and if so, are they ultimately a consequence of the symmetries of the inverted repeats? It is tempting to speculate that occluded states are the fully symmetrical high energy conformations whose energy is lowered by symmetry breaking. The simple unified model indicates that transporter function may form a continuum. However, how difficult is it to move through this continuum and what are the minimal changes to change physiological function? How can such changes be achieved with allosteric modulators (small molecules) or changes in external conditions such as membrane tension, pressure, temperature, or transmembrane voltage?

More broadly speaking, it has also been recognized that channels and transporters form a spectrum \citep{Ashcroft:2009vn, Gadsby:2009uq, Henderson:2019aa}, or as expressed by \citet{Lauger:1980qf}: ``Channel and carrier [transporter] models should therefore not be regarded as mutually exclusive possibilities, but rather as limiting cases of a more general mechanism.'' There seems to be value in stepping back and asking what the general principles are under which a class of proteins have to operate.

\begin{acknowledgments}
  We thank Philip Fowler for providing us with the data for Figure~\ref{fig:mfsconformations}.  Research reported in this work was supported by the National Institute Of General Medical Sciences of the National Institutes of Health under Award No R01GM118772. Some of the work was supported by the European Community’s Seventh Framework Programme FP7/2007–2013 under Grant Agreement No HEALTH-F4-2007-201924, European Drug Initiative for Channels and Transporters (EDICT) Consortium.

Some molecular graphics were produced with Visual Molecular Dynamics (VMD, \url{ http://www.ks.uiuc.edu/Research/vmd/}). VMD was developed by the Theoretical and Computational Biophysics Group in the Beckman Institute for Advanced Science and Technology at the University of Illinois at Urbana-Champaign. Some molecular graphics were produced with UCSF Chimera, developed by the Resource for Biocomputing, Visualization, and Informatics at the University of California, San Francisco, with support from NIH P41-GM103311. Some molecular graphics were produced with the PyMOL Molecular Graphics System (Schr{\"o}dinger, LLC).
\end{acknowledgments}

\section*{Author declarations}
\paragraph*{Conflict of Interest} The authors have no conflict of interest to disclose.

\section*{Data availability}
The data that support the findings of this study are available from the corresponding author upon reasonable request.

\bibliography{journals, becksteinlab}

\end{document}